\pdfoutput=1
\documentclass[prd,aps,nofootinbib,twocolumn,superscriptaddress,preprintnumbers,balancelastpage,longbibliography]{revtex4-1}
\usepackage{ae,aecompl}
\usepackage{graphicx}	
\usepackage{amsmath}	
\usepackage{graphicx}
\usepackage{dcolumn}
\usepackage{bm}
\usepackage{graphics}
\usepackage{afterpage}
\usepackage{float}
\usepackage{subfigure}
\usepackage{rotating}
\usepackage{multirow}
\usepackage{tabularx}
\usepackage{booktabs}
\usepackage{multirow}
\usepackage{fancyhdr}
\usepackage{hyperref}
\usepackage[utf8]{inputenc}
\usepackage{theorem}
\usepackage{moreverb}
\usepackage{euscript}
\usepackage{psfrag}
\usepackage{slashed}
\usepackage{mathtools}
\usepackage{makecell}
\usepackage[flushleft]{threeparttable}
\usepackage{adjustbox}
\usepackage[english]{babel}

\usepackage{dcolumn}
\usepackage{bm}
\usepackage[mathlines]{lineno}

\hypersetup{
     colorlinks   = true,
     citecolor    = magenta,
     urlcolor     = magenta,
     linkcolor    = magenta
}

\usepackage{listings}
\usepackage{color,xcolor}

\newcolumntype{C}[1]{>{\centering\let\newline\\\arraybackslash\hspace{0pt}}m{#1}}

\lstdefinestyle{python}{
  belowcaptionskip=1\baselineskip,
  breaklines=true,
  frame=L,
  xleftmargin=\parindent,
  language=Python,
  showstringspaces=false,
  basicstyle=\small\ttfamily,
  morekeywords={models, lambda, forms,True,False,None},
  keywordstyle=\bfseries\color{deepgreen!40!black},
  commentstyle=\itshape\color{gray},
  identifierstyle=\color{black},
  stringstyle=\color{deepred},
  rulecolor=\color{gray},
}

\begin{document}

\preprint{MIT-CTP/5230}
\preprint{SLAC-PUB-17556}

\title{Exoplanets as Sub-GeV Dark Matter Detectors}

\author{Rebecca K. Leane}
\thanks{{\scriptsize Email}: \href{mailto:rleane@slac.stanford.edu}{rleane@slac.stanford.edu}; {\scriptsize ORCID}: \href{http://orcid.org/0000-0002-1287-8780}{0000-0002-1287-8780}}
\affiliation{Center for Theoretical Physics, Massachusetts Institute of Technology, Cambridge, MA 02139, USA}
\affiliation{SLAC National Accelerator Laboratory, Stanford University, Stanford, CA 94039, USA}

\author{Juri Smirnov}
\thanks{{\scriptsize Email}: \href{mailto:smirnov.9@osu.edu}{smirnov.9@osu.edu}; {\scriptsize ORCID}: \href{http://orcid.org/0000-0002-3082-0929}{0000-0002-3082-0929}}
\affiliation{Center for Cosmology and AstroParticle Physics (CCAPP), The Ohio State University, Columbus, OH 43210, USA}
\affiliation{Department of Physics, The Ohio State University, Columbus, OH 43210, USA}

\date{\today}

\newcommand{\mk}[1]{{\bf #1}}
\newcommand{\om}[1]{\textcolor{red}{#1}}
\newcommand{\sh}[1]{\textcolor{blue}{#1}}
\newcommand{\aap}{Astronomy and Astrophysics}
\newcommand{\mnras}{Monthly Notices of the RAS}

\begin{abstract}
We present exoplanets as new targets to discover Dark Matter (DM). Throughout the Milky Way, DM can scatter, become captured, deposit annihilation energy, and increase the heat flow within exoplanets. We estimate upcoming infrared telescope sensitivity to this scenario, finding actionable discovery or exclusion searches. We find that DM with masses above about an MeV can be probed with exoplanets at DM-proton and DM-electron scattering cross sections down to about $10^{-37}$cm$^2$, stronger than existing limits by up to six orders of magnitude. Supporting evidence of a DM origin can be identified through DM-induced exoplanet heating correlated with Galactic position, and hence DM density. This provides new motivation to measure the temperature of the billions of brown dwarfs, rogue planets, and gas giants peppered throughout our Galaxy.
\end{abstract}

\maketitle

\noindent\textbf{\textit{Introduction--}}Are we alone in the Universe? This question has driven wide-reaching interest in discovering a planet like our own. Regardless of whether or not we ever find alien life, the scientific advances from finding and understanding other planets will be enormous. From a particle physics perspective, new celestial bodies provide a vast playground to discover new physics. 

Astrophysical systems have already been broadly used to probe new physics, including investigating the effects of gravitationally captured Dark Matter (DM). If there is sufficient gravitational force, deposited DM kinetic energy can noticeably increase the temperature of the system. Regardless of gravitational strength, DM annihilation can also induce heating. This has been investigated in the context of neutron stars and white dwarfs~\cite{
Goldman:1989nd,
Gould:1989gw,
Kouvaris:2007ay,
Bertone:2007ae,
deLavallaz:2010wp,
Kouvaris:2010vv,
McDermott:2011jp,
Kouvaris:2011fi,
Guver:2012ba,
Bramante:2013hn,
Bell:2013xk,
Bramante:2013nma,
Bertoni:2013bsa,
Kouvaris:2010jy,
McCullough:2010ai,
Perez-Garcia:2014dra,
Bramante:2015cua,
Graham:2015apa,
Cermeno:2016olb,
Graham:2018efk,
Acevedo:2019gre,
Janish:2019nkk,
Krall:2017xij,
McKeen:2018xwc,
Baryakhtar:2017dbj,
Raj:2017wrv,
Bell:2018pkk,
Garani:2018kkd,
Chen:2018ohx,
Garani:2018kkd,
Dasgupta:2019juq,
Hamaguchi:2019oev,
Camargo:2019wou,
Bell:2019pyc,
Garani:2019fpa,
Acevedo:2019agu,
Joglekar:2019vzy,
Joglekar:2020liw,
Bell:2020jou,
Dasgupta:2020dik,
Garani:2020wge,
Leane:2021ihh}. Alternatively, the DM-related heat flow in other moons and planets has been considered, including Earth~\cite{Mack:2007xj,Chauhan:2016joa,Bramante:2019fhi}, Uranus~\cite{Mitra:2004fh,Adler:2008ky}, Neptune and Jupiter~\cite{Adler:2008ky,Kawasaki:1991eu}, Mars~\cite{Bramante:2019fhi}, Earth's Luna~\cite{Garani:2019rcb,Chan:2020vsr}, as well as hot Jupiters~\cite{Adler:2008ky}.

\begin{figure}[t!]
\centering
\includegraphics[width=\columnwidth]{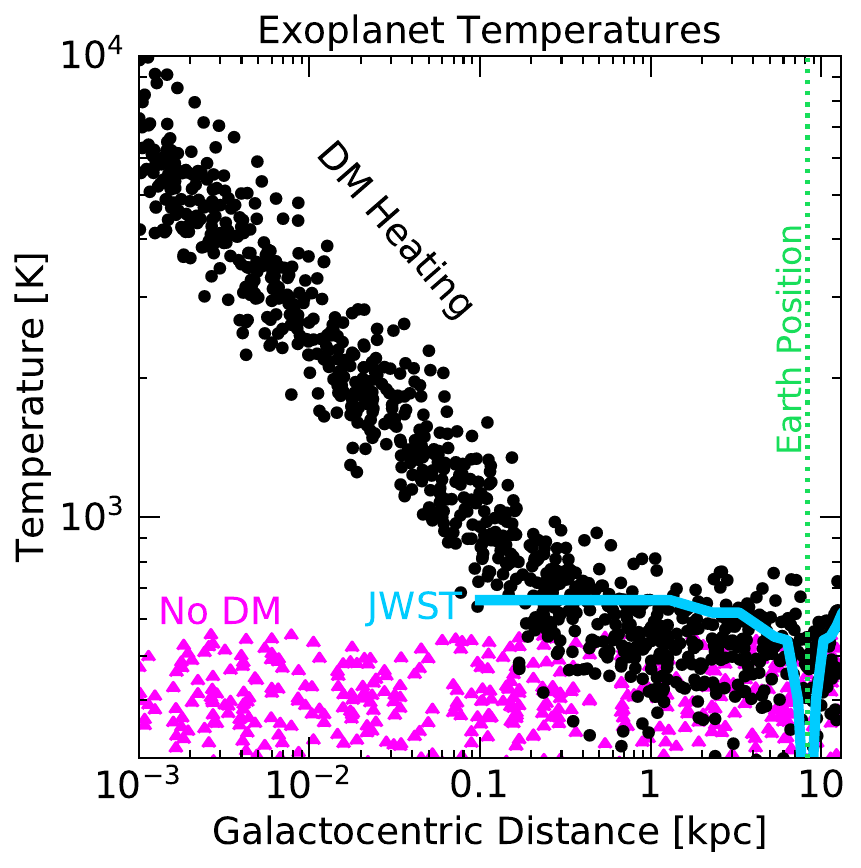}
\caption{Mock temperature distribution of old example exoplanets with $20-50$ Jupiter masses over Galactocentric distances. Black dots are DM-heated exoplanets, magenta triangles are the same set of planets, without DM heating. JWST is the estimated minimum telescope sensitivity (see text).}
\label{fig:schem}
\end{figure}

We explore the potential to discover DM using \textit{exoplanets} -- planets outside our solar system. We will use the term ``exoplanets'' to refer to the broader class of all extra-solar planets (including rogue planets), as well as brown dwarfs, which exist at the planet-star boundary. The general setup of this idea is as follows: particle DM in the galactic halo can scatter with exoplanets, lose energy, and become gravitationally captured by the exoplanet. The captured DM accumulates and may annihilate, releasing its mass energy to heat exoplanets. Assuming that the annihilation rate is in equilibrium with the scattering rate (see Supplementary Material), the annihilation heat measured by upcoming infrared telescopes allows for a new probe of the DM scattering rate. 

We will show that this leads to new sensitivities to scattering cross sections between about $10^{-37} - 10^{-25}$~cm$^2$ in the sub-GeV mass range. This range of elastic interactions is expected in models with thermally produced sub-GeV DM, see e.g. Refs.~\cite{Hochberg:2014dra,Smirnov:2020zwf}. This cross section range is bounded by sufficiently weak DM interactions for DM to drift to the core and accumulate, and sufficiently strong DM interactions to produce a detectable DM heat flux. This requires the annihilation rates to be larger than a lower bound provided by the capture and annihilation equilibration condition. For e.g. $2 \rightarrow 2$ annihilation, we will show that the thermally averaged cross section must be greater than about $10^{-37} - 10^{-34}\,\text{cm}^3/\rm s$ depending on the target, such that both $s-$ and $p-$wave annihilation can be probed (see Supplementary Material for more details). The lower DM-mass sensitivity is truncated by DM evaporating from the exoplanet (and therefore not annihilating to produce any heat), with sensitivity depending on the DM model. While higher DM masses can also be probed with exoplanets, we will focus on the MeV$-$GeV mass range, as this features a new cross section range that has not been probed by direct detection or other experiments.

There are many advantages of using exoplanets to search for DM over other celestial bodies. These include:

\textbf{A rapidly accelerating research program:} Until 1992, we didn't even know if exoplanets existed. Almost all exoplanets we now know were only discovered in the last decade, with the majority found in the last five years~\cite{catalog}. The exoplanet program is clearly rapidly growing (see Supplementary Material for details of many new experiments). This provides ample motivation to consider new ways this exploding research area can be used to probe new physics.

\textbf{Enormous number of expected exoplanets:} It is estimated that there is at least one planet per star in our Galaxy, and about one cold planet per star~\cite{Cassan_2012}. This means that there should be about 300 billion exoplanets awaiting discovery. While of course these won't all be immediately found, even a small percentage of this number leads to an enormous statistical advantage for understanding potential signals.  
It also allows ample room for growth with new discoveries and possible surprises in observations. To date, there are 4,324 confirmed exoplanets, and an additional 5,695 candidates are currently under investigation~\cite{catalog}.

\textbf{Much larger surface area than neutron stars:} The other key proposed search using upcoming infrared telescopes on DM-heated astrophysical bodies is with old, cold neutron stars~\cite{Baryakhtar:2017dbj}. However, while neutron stars are much more dense, and allow for higher heating rates in part due to enhancements from kinetic heating, exoplanets and brown dwarfs are \textit{much} larger. A typical neutron star has a radius of about~\mbox{10 km}, while exoplanets of interest have radii of about 50,000 -- 200,000 km. This means that exoplanet temperatures can be measured much further into the GC and therefore can provide a DM-density dependent heating signal. Exoplanets can also be imaged to much higher significance, and with less exposure time.

\textbf{Easier to find than neutron stars:} The infrared neutron star search requires that a sufficiently cold neutron star candidate at a distance $\lesssim$ 100 pc from Earth is found~\cite{Baryakhtar:2017dbj}. While pulsars have been found at distances of $\sim 100~{\rm pc}$~\cite{Manchester:2004bp}, it is possible that a sufficiently cold and sufficiently close-by neutron star may not be found, or cannot be measured with sufficient exposure time. On the other hand, exoplanets outnumber neutron stars in our galaxy by at least about a factor of a thousand~\cite{Camenzind}, and are already known to exist in close enough proximity for DM searches.

\textbf{Low temperatures:} Lastly, exoplanets can be very cold, as they do not undergo nuclear fusion, and can exist very far in large orbits from any host star to which they may be bound. They can even go \textit{rogue}, floating free from any parent star. As the low temperatures allow for a clearer signal over background for DM heating, exoplanets are advantageous over nuclear-fusing stars. Furthermore, their low core temperatures in part prevent DM evaporation compared to evaporation in these stars, providing new sensitivity to MeV DM.

In this work, we exploit all these features to identify new searches for DM in exoplanets. We establish two different searches: one for distant exoplanets and one for local exoplanets. The distant exoplanet searches will require that the exoplanets be rogue (or brown dwarfs), such that their detection is not obscured.

Figure~\ref{fig:schem} demonstrates these searches and shows an example distribution of exoplanets with masses of about $20-50$ Jupiters, with and without DM heating. Distant exoplanets can be used to map the Galactic DM density, given sufficient telescope sensitivity. This is seen by the uptick of many hot exoplanets, scaling with the DM density. As well as searching for DM signals, local exoplanets can be used to test the hypothesis that DM contributes to internal heat of the gas giants in our own solar system, which are not well understood~\cite{Kawasaki:1991eu,Adler:2008ky}. DM-heated exoplanets can be potentially measured when the infrared telescope JWST comes online. Both our suggested searches target new DM parameter space, probing 
the DM-proton and DM-electron scattering cross sections to unprecedented sensitivities.\\

\noindent\textbf{\textit{Dark Heat Flow in Exoplanets--}}
The total heat flow of the exoplanet $\Gamma_{\rm heat}^{\rm tot}$ can be determined by combining potential heat power sources, including internal heat $\Gamma_{\rm heat}^{\rm int}$, external heat $\Gamma_{\rm heat}^{\rm ext}$, and DM heat $\Gamma_{\rm heat}^{\rm DM}$:
\begin{equation}
 \Gamma_{\rm heat}^{\rm tot}=\Gamma_{\rm heat}^{\rm ext}+\Gamma_{\rm heat}^{\rm int}+\Gamma_{\rm heat}^{\rm DM}= 4\pi R^2 \, \sigma_{\rm SB} \, T^4 \, \epsilon,
 \label{eq:heat}
\end{equation}
where $R$ is the exoplanet radius, $T$ is the exoplanet temperature, $\sigma_{\rm SB}$ is the Stefan-Boltzmann constant, and $\epsilon$ is the emissivity (which is a measure of planetary heat radiation efficiency, ranging from 0 to 1). External heating is negligible for wide-orbit or free-floating planets.

\begin{figure*}[t!]
\centering
\includegraphics[width=0.683\columnwidth]{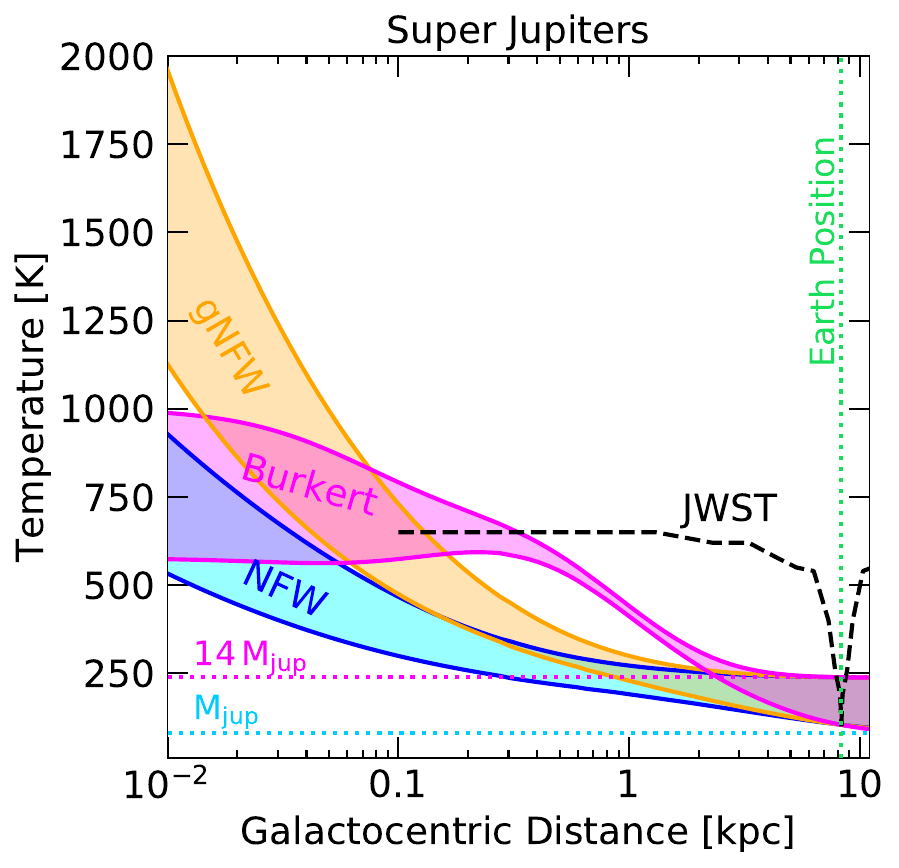}
\includegraphics[width=0.683\columnwidth]{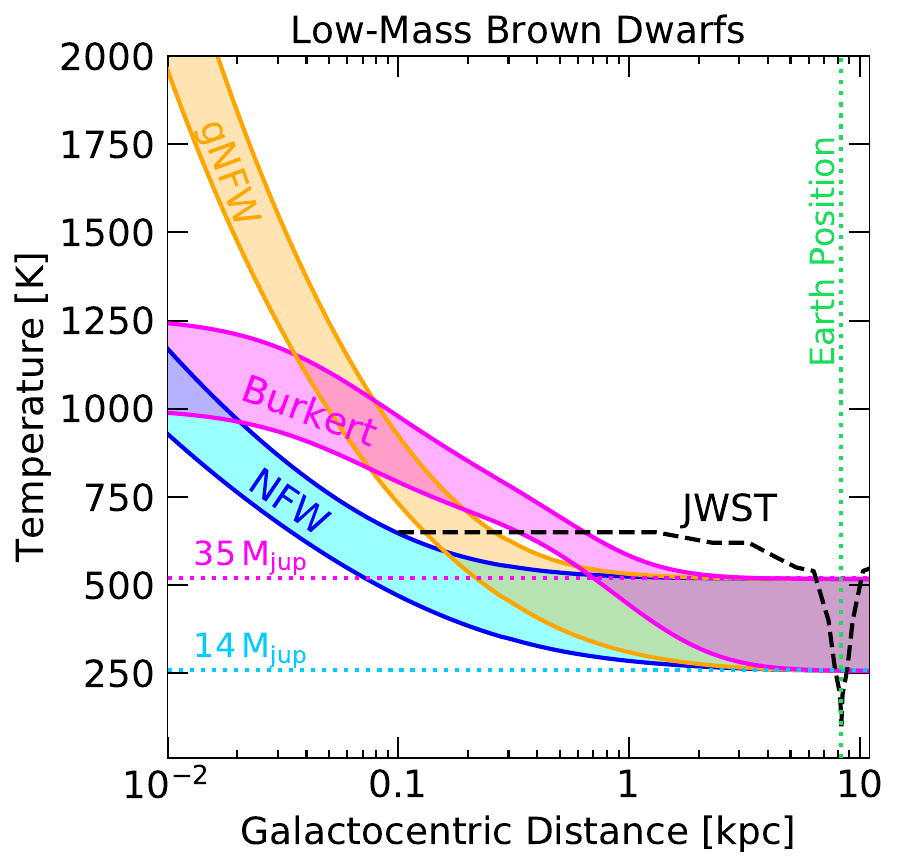}
\includegraphics[width=0.683\columnwidth]{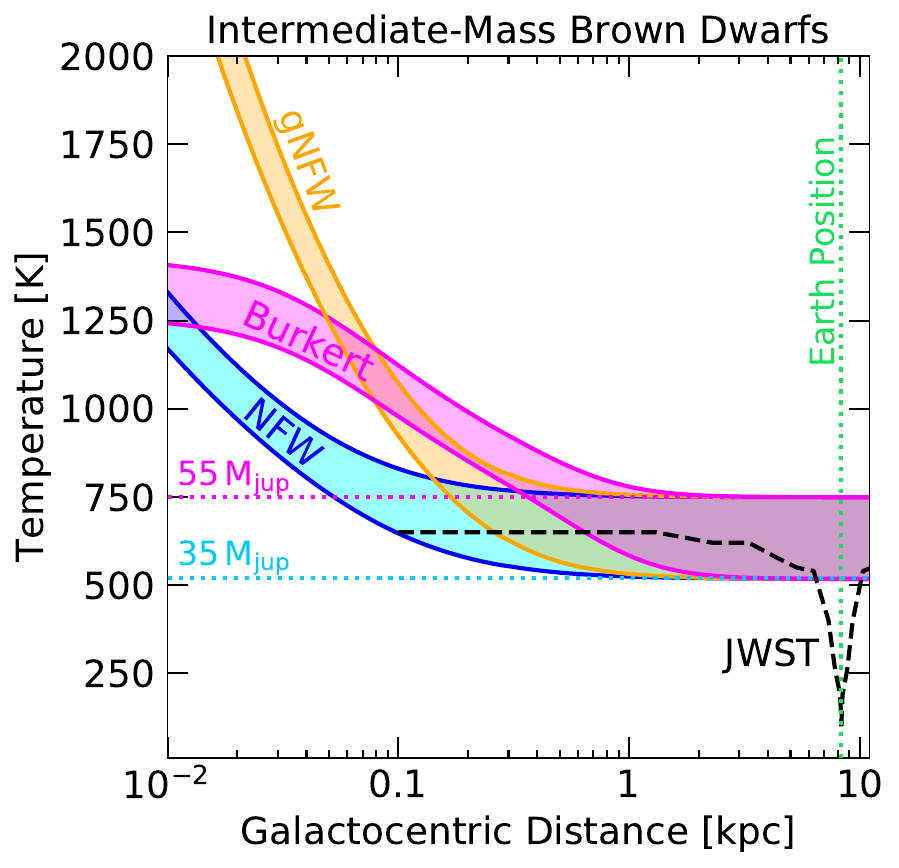}
\caption{Exoplanet temperatures over Galactocentric distances, with variations due to DM for labeled density profiles. Each panel represents our classification of different exoplanet types: Super Jupiters ($M_{\rm jup}-14\,M_{\rm jup}$), low-mass brown dwarfs ($14\,M_{\rm jup}-35\,M_{\rm jup}$), and intermediate-mass brown dwarfs ($35\,M_{\rm jup}-55\,M_{\rm jup}$). Any exoplanet within the indicated mass range will have temperatures between these lines in the shaded region. The dotted lines show the range of minimum temperatures for a 10 Giga-year-old exoplanet without DM. Black dashed line is the optimal minimum JWST sensitivity (temperatures above this line can be detected), see text.}
\label{fig:profiles}
\end{figure*}

We compute the internal heat flow for our range of benchmark brown dwarfs and Jupiters without DM. As the minimum temperature for heavy brown dwarfs (with $55\,M_{\rm jup}$) and benchmark Jupiters (with $M_{\rm jup}$) after about 10 Gyr is about 750~K and 80~K respectively, we can determine the internal heat flow required to produce these temperatures,
\begin{equation}
 \Gamma_{\rm heat}^{\rm int}=4\pi R^2 \, \sigma_{\rm SB} \, T^4\, \epsilon,
\end{equation}
which corresponds to about $1.1\times10^9$ TW and $1.4\times10^5$ TW for benchmark brown dwarfs and Jupiters, respectively. This serves as our non-DM baseline for comparing with a potential DM signal.

The additional DM heating source occurs if DM scatters on exoplanet particles, becomes captured, and annihilates. This produces heat that can be absorbed by the exoplanet. We assume that the DM scattering and annihilation processes are in equilibrium (see Supplementary Material). The DM heat flow depends on how many external DM particles are captured from the incoming DM flux reservoir. The maximal, i.e. geometric capture rate of DM is given by~\cite{Garani:2017jcj}
\begin{equation}
C_{\rm max} = \pi\,R^2 \,n_{\chi}(r)\,v_0 \left( 1 + \frac{3}{2}\frac{v_{\rm esc}^2}{v_{d}(r)^2} \right) \,\xi( v_{\rm p},v_{d}(r)) \,,
\label{eq:geom}
\end{equation}
where $n_\chi(r)$ is the DM number density at distance $r$ from the GC, the average speed in the DM rest frame $ v_0$ is related to the velocity dispersion $v_{d}(r)$ as $v_0= \sqrt{8/(3\,\pi)} v_{d}(r)$ at distance $r$ from the GC, and $R$ is the exoplanet radius. The factor $1 +  3\,v_{\rm esc}^2/2v_d^2$ is the result of gravitational focusing, with $v_{\rm esc}^2 = 2 G_N M/R$ being the escape velocity, $M$ the exoplanet mass, and $G_N$ the gravitational constant. The motion of the planet with velocity $v_p$ with respect to the DM halo is taken into account by $\xi(v_{\rm p},v_{d}(r))$. In the scenarios we are interested in, the DM velocity, the planetary velocity and the escape velocities are of similar order and the function $\xi( v_{d}(r), v_{\rm p}) \sim 1$. The circular velocities $v_c(r)$ in the galaxy are related to the DM velocity dispersion by $v_{d}(r) = \sqrt{3/2} \,v_c(r)$. We extract the circular velocities at different radii in the Milky Way by combining the data for the gas, bulge, and disk components, as well as the analytic expressions for DM contributions to the total velocity from Ref.~\cite{Lin:2019yux}. For the DM density, we consider an Navarro-Frenk-White (NFW) profile~\cite{Navarro_1996}, a generalized NFW (gNFW) profile, and a Burkert profile~\cite{Burkert_1995}, with the local DM density $0.42$~GeV/cm$^3$~\cite{Pato_2015}, see Supplementary Material for more details. Eq.~\ref{eq:geom} does not include reflection, which can weaken sensitivities when DM is light and the object has a low escape velocity compared to the DM halo velocity~\cite{Neufeld:2018slx,Leane:2022hkk}, depending on the DM model interpretation.

The heat power produced by DM is given by the product of the DM mass $m_\chi$, the fraction of the captured DM particles that have passed through the object $f$, and the maximal capture rate, such that
\begin{equation}
 \Gamma_{\rm heat}^{\rm DM}= m_{\chi}\,f\, C_{\rm max}.
\end{equation}
Using $n_\chi(r)=\rho_\chi(r)/m_\chi$, $\xi( v_{d}(r), v_{\rm p}) \sim 1$, and combining with Eq.~\ref{eq:geom}, the DM heat power is
\begin{equation}
 \Gamma_{\rm heat}^{\rm DM}= f\, \pi R^2\rho_{\chi}(r)\, v_0 \left( 1+\frac{3}{2}\frac{v_{\rm esc}^2}{v_d(r)^2} \right).
\end{equation}

\noindent\textbf{\textit{Searches and Infrared Telescope Sensitivity--}}Exoplanets may first be identified by e.g. Doppler spectroscopy or gravitational lensing. Once their location is found, infrared telescopes such as JWST may be able to measure their temperature. The general sensitivity of JWST to exoplanet heating can be found by considering the spectral flux density,
\begin{equation}
 f_\nu=\pi B(\nu,T)\times\frac{4\pi R^2}{4\pi d^2},
 \label{eq:flux}
\end{equation}
where $d$ is the distance from the telescope to the exoplanet, $R$ is the radius of the exoplanet, and
\begin{equation}
 B(\nu,T)=\frac{2\nu^3\epsilon}{{\rm exp\left(\frac{2\pi\nu}{k_b T}\right)-1}},
\end{equation}
where $\nu$ is the frequency, $T$ is the temperature, $k_b$ is the Boltzmann constant, and $\epsilon$ is the atmospheric emissivity. We use $\epsilon=1$ which provides the usual blackbody spectral flux density, and is the most conservative case. Deviations from a blackbody occur for \mbox{$\epsilon<1$}; see Supplementary Material for emissivity impact on telescope sensitivity.

Figure~\ref{fig:profiles} shows the expected exoplanet temperature as a function of Galactocentric distance, for DM-heating arising due to several different DM profiles as labeled. We distinguish between Jovian exoplanets with masses between $1- 14\,M_{\rm jup}$ and brown dwarfs with masses in the range of $14 -55\,M_{\rm jup}$.  All exoplanets shown have a radius of $R_{\rm jup}$, as they are expected to roughly converge to this radius after 10 Gy.  The shaded region for a given DM profile represents the range of heating possibilities for the indicated mass range, with the heaviest (lightest) exoplanets lying at the upper (lower) boundary. The shape of the curves over galactic distances is due to an interplay of the DM density and velocity profiles, and the effective capture radius of the exoplanet.

We show in Fig.~\ref{fig:profiles} the optimal JWST sensitivity, which is found using Eq.~\ref{eq:flux} with the benchmark dwarf/Jupiter radius of $R_{\rm jup}$. As different JWST instrument filters are optimized for different flux densities/temperatures, we use several different filters while scanning over the minimal temperature measurable, to obtain the optimal sensitivity. This is calculated using the \textsc{Near-Infrared Imager and Slitless Spectrograph (NIRISS)} in \textsc{Imaging} mode for temperatures above about 500~K, and the \textsc{Mid-Infrared Instrument (MIRI)} in \textsc{Imaging} or \textsc{Medium-Resolution Spectroscopy} mode for temperatures from about $100-500$ K. The dashed line is for JWST to obtain about $10^5$ seconds of exposure to achieve a signal-to-noise ratio (SNR) of 2. 10 SNR can be achieved at about $10^6$ seconds of exposure at most of the temperatures shown. Note however that these exposure times are for the minimum temperatures on the dashed line; higher temperatures generally require less exposure time. Significantly less time is required to achieve 10 SNR in the local region; see Supplementary Material for more detailed JWST sensitivity estimates, including estimates of dust extinction and stellar crowding, which conservatively limit the expected JWST sensitivity to distances larger than 0.1 kpc from the GC.

The dotted lines in Fig.~\ref{fig:profiles} show exoplanet temperatures without DM heating, for masses labeled, where intermediate masses would sit between these dotted lines. It is clear that different types of exoplanets are most useful as DM heating targets in different regimes. The lower mass Jupiters are ideal for local searches, as DM heating can outperform their internal heat at the local position. For higher mass brown dwarfs, their internal heat is too high to reveal a DM heating signal at the local position. However, their larger masses are advantageous towards the GC due to gravitational focusing. 

 We therefore propose two search strategies: one for DM in local Jupiters, and another for all exoplanets towards the DM dense GC. As shown in Fig.~\ref{fig:profiles}, towards the GC exoplanets are increasingly heated by DM. This means that exoplanets can potentially be used to trace the DM density in our Galaxy. Given a large statistical sample, DM overdensities could also be revealed by too many hot exoplanets in a given region, which is additional motivation for the distant search.
Note that at large distances the exoplanet must be a rogue planet or brown dwarf, since a parent star would obscure its resolution.

\vspace{4mm}

\noindent\textbf{\textit{Dark Matter Parameter Space--}}We now consider the implications of DM-heated exoplanets for particle DM models. Limits from planetary heat flow often probe strongly interacting DM (SIMPs)~\cite{Mack:2007xj,Bramante:2019fhi}. Exoplanets are instead ideal laboratories to study broader classes of sub-GeV DM models. 

To relate the DM heat flow with scattering cross sections, we need to find the range of parameters for which a fraction $f$ of the DM particles passing through the planet is gravitationally captured. 
Normalizing to the maximal DM capture rate (defined in Eq.~\ref{eq:geom}), we obtain the captured DM fraction 
\begin{align}
\label{eq:capturecondition}
f = \frac{C_{\rm cap} }{C_{\rm max}} =  \sum_{N=1}^{N_{\rm max}}  \,f_N\,,
\end{align}
with the capture fraction for a given number of scatterings being
\begin{eqnarray}
f_N & = & p(N,\tau)  \left[ 1 -
\kappa \exp{\left( -\frac{3  \left(v_{\rm N}^2 - v_{\rm esc}^2 \right)}{2 v_d^2}\right) } \right],
\end{eqnarray}
with 
\begin{align}
\kappa = \left(1 + \frac{3}{2}\frac{ v_{\rm N}^2}{v_d^2} \right)  \left( 1 + \frac{3}{2} \frac{v_{\rm esc}^2}{ v_d^2}\right)^{-1}\,.
\end{align}
Here $v_d$ is the velocity dispersion, $v_N = v_{\rm esc} \left(1 - \langle z\rangle\beta\right)^{-N/2}$ where the average scattering angle is $\langle z\rangle=1/2$~\cite{Bramante:2017xlb,PhysRevD.102.048301}, $\beta = 4 m_{\chi} m_A/(m_\chi + m_A)^2$, and $m_A$ is the mass of the target particle. The probability that the DM particle scatters $N$ times is~\cite{Bramante:2017xlb,PhysRevD.102.048301}
\begin{equation}
p(N,\tau) =  \frac{2}{\tau^2} \left( N_s +1  - \frac{\Gamma(N_s +2, \tau)}{N_s!} \right) \, ,
\end{equation}
where $\Gamma(a,b)$ is the incomplete gamma function. This scattering probability is a function of the optical depth, $\tau = \frac{3}{2} \,\sigma/\sigma_{\rm sat} $ where $\sigma_{\rm sat} = \pi R^2/N_{\rm SM}$ is the saturation cross section, $R$ the planetary radius, $N_{\rm SM}$ is the target particle number, and $\sigma$ is the DM-target cross section. To set sensitivity limits on DM scattering in Jupiters and brown dwarfs, we assume spheres of hydrogen with constant density. As gas giants are expected to be dominated by hydrogen, we expect a hydrogen sphere to be approximately representative.  For example, Jupiter's composition is about $84 \, \%$ hydrogen, and $16\, \%$ helium~\cite{Guillot:2009ij}. 
 
Figure~\ref{fig:xsec} shows our sensitivity estimates for Jupiter-like planets and brown dwarfs to the DM parameter space for spin-dependent DM-proton scattering (see Sec.~V of the Supplementary Material for spin-independent, additional velocity dependent, and electron scattering results). 
We show a ``min'' cross section for the object as labelled, which corresponds to effectively all the DM being captured (about $95\%$), and planets being maximally heated, producing the most striking signal. We also show a ``max'' cross section, which corresponds to the smallest DM capture fraction (about 10$\%$) that can be probed in the near future with JWST. This is likely the maximum cross section reach, as the temperatures corresponding to this lower scattering rate are approaching either the JWST minimum temperature detection threshold, or the expected background temperature, in most of the parameter space. The sensitivity curves become flat for brown dwarfs, as their larger escape velocity leads to the opacity limit being reached for some of the parameter space. Thus, increasing the DM capture cross section further in that region does not change the capture fraction. The lower end of the curves is truncated by the mass in which DM evaporates out of the system. Here, we have not included evaporation as this value is highly model dependent; see Supplementary Material for discussion.

\begin{figure}[t!]
\centering
\includegraphics[width=\columnwidth]{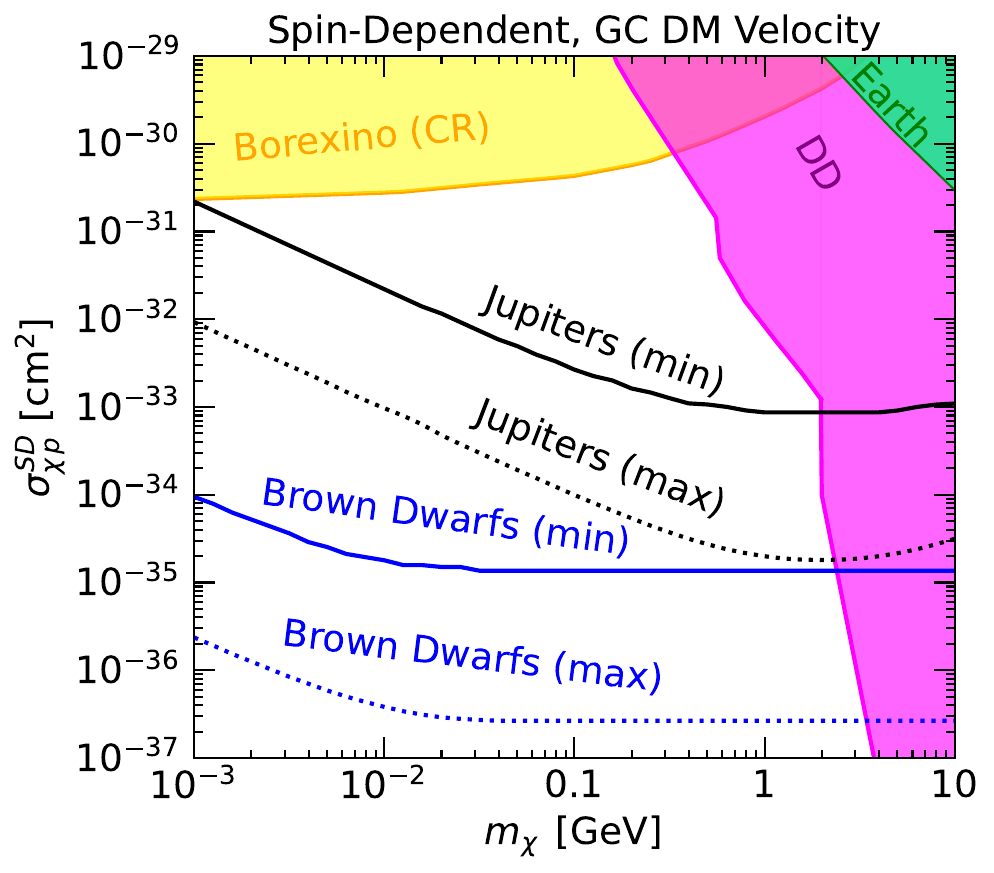}
\caption{Spin-dependent DM-nucleon scattering cross section sensitivity estimates for Jupiters and brown dwarfs, assuming a GC DM velocity. The solid (min) lines show cross sections assuming effectively all DM is captured, and the dotted (max) lines show the maximum expected reach. Complementary constraints are also shown, see text.}
\label{fig:xsec}
\end{figure}

These sensitivities do not depend on the DM density profile or value; sensitivity to the cross sections shown only requires that the DM-heating is successfully detected. For more discussion of detection/search prospects, see the Supplementary Material. We show the earth heat flow bounds from Ref.~\cite{Bramante:2019fhi} for comparison, and direct detection (DD) bounds from DM-proton scattering~\cite{Gangi:2019zib, Aprile:2019dbj, Aprile:2019jmx,Liu:2019kzq}. Borexino(CR) shows limits from boosted DM from cosmic rays interactions~\cite{Cappiello:2018hsu, Bringmann:2018cvk,Ema:2018bih, Cappiello:2019qsw}. Note that these limits have different assumptions; the direct detection limits do not require any minimum annihilation cross section.

\vspace{4mm}

\noindent\textbf{\textit{Conclusions--}}The exoplanet program is rapidly accelerating. Amongst the billions of new worlds in our Galaxy, many are waiting to reveal their surprises. Unexpected discoveries are inevitable, and numerous new telescopes with cutting-edge technology are ready to make them. For the first time, we have pointed out the broad applicability of exoplanets to be used as DM detectors, with actionable discovery or exclusion searches using new infrared telescopes. We target old, cold, Jupiter-like planets and brown dwarfs, which are advantageous due to their large sizes, densities, and low core temperatures.

Our first suggested search is for overheated local exoplanets. There are hundreds of known Jupiters in our local neighborhood, and Gaia is expected to identify tens of thousands of potential candidates in the next few years~\cite{Perryman_2014}, providing a sample with great statistical power.

Our second suggested search is for overheated exoplanets, correlated with DM density, rising sharply in temperature towards the GC. The presence of DM overdensities or substructure may also be confirmed with exoplanets, with a pocket of even hotter DM-heated exoplanets. We conclude that, at an estimate, JWST may have sensitivity to exoplanet temperatures above about 650 K, for exoplanets all the way into about 0.1 kpc of the inner Galaxy (for more local searches, the minimum temperature sensitivity is lower).

We calculated the DM parameter space sensitivity to brown dwarfs and Jupiters that may have their temperature measured in the near future. We determined that DM with masses above about an MeV can be probed with exoplanets at DM-proton scattering cross sections down to about $10^{-37}~$cm$^2$, stronger than existing limits by up to six orders of magnitude. We pointed out that this DM mass sensitivity is lighter than many other celestial body searches for DM heat flow. This is because brown dwarfs and Jupiters have large integrated column densities and low core temperatures, and so it is more difficult for light DM to evaporate in these systems.

For further details on exoplanets, particle DM models, additional search details and cross section sensitivities, see the Supplementary Material.\\

\noindent\textbf{\textit{Acknowledgments--}}We thank John Beacom, Joe Bramante, Chris Cappiello, Rouven Essig, Katie Freese, Raghuveer Garani, Savannah Jacklin, Shirley Li, Bruce Macintosh, Nirmal Raj, Anupam Ray, Pat Scott, Sara Seager, Aaron Vincent, and Ji Wang for helpful comments and discussions. RKL was supported by the Office of High Energy Physics of the U.S. Department of Energy under Grant No. DE-SC00012567 and DE-SC0013999, as well as the NASA Fermi Guest Investigator Program under Grant No. 80NSSC19K1515, and later at SLAC under Contract DE-AC02-76SF00515. JS is largely supported by a Feodor Lynen Fellowship from the Alexander von Humboldt foundation.

\clearpage
\newpage
\maketitle
\onecolumngrid
\begin{center}
\textbf{\large Exoplanets as Sub-GeV Dark Matter Detectors}

\vspace{0.05in}
{ \it \large Supplementary Material}\\ 
\vspace{0.05in}
{Rebecca K. Leane and Juri Smirnov}
\end{center}
\onecolumngrid
\setcounter{equation}{0}
\setcounter{figure}{0}
\setcounter{section}{0}
\setcounter{table}{0}
\setcounter{page}{1}
\makeatletter
\renewcommand{\theequation}{S\arabic{equation}}
\renewcommand{\thefigure}{S\arabic{figure}}
\renewcommand{\thetable}{S\arabic{table}}

The Supplementary Material contains additional details relevant to our searches, as detailed in the Table of Contents below. The most important contents of the Supplementary Material are the additional cross section sensitivity results, which include spin-dependent scattering in local DM velocities in Fig.~\ref{fig:spindeploc}, spin-independent scattering in Fig.~\ref{fig:spinindep}, and electron scattering results in Fig.~\ref{fig:xseclep}.

\tableofcontents

\newpage
\section{Overview of Exoplanets}
\label{sec:overview}
In this section we give a brief overview of the available menu of exoplanets and argue that large gaseous planets, such as super-Jupiters and brown dwarfs, are advantageous for our DM search strategy.

\subsection{Abundance and Properties}

It is expected that on average, all stars have at least one planet~\cite{Cassan_2012}. Given we know that there are about 300 billion stars in our Galaxy, this amounts to about 300 billion or more exoplanets in the Milky Way. Of these, there is a smorgasbord of exoplanet types, with diverse properties and sizes, which we briefly outline below. 
\vspace{-4mm}

\subsubsection{Earth-like Planets}
The most popular exoplanet type for finding aliens are, of course, Earth-like planets. Earth-like planets have rocky interiors, and relatively small masses and radii relative to all other exoplanets. They extend into the ``Super-Earths'' category, which usually have radii of about that of Earth, but have up to a factor 10 higher in mass. These are not ideal for our DM searches, as their radii are smaller than other exoplanets, leading to limitations in telescope sensitivity. Interestingly however, it has been pointed out that DM annihilation heating of Earth-like exoplanets can lead to liquid water, and therefore a habitable planet, when otherwise the planet would have been too cold~\cite{Hooper:2011dw}.  
\vspace{-6mm}

\subsubsection{Jupiters}
The next-largest category is the gas giants, also called ``Jovian planets'' or ``Jupiters''. Jupiters have radii roughly comparable to that of Jupiter, and generally have masses about comparable to Jupiter, though they can have up to about 10 times higher masses, becoming ``Super Jupiters'' (any higher, and they begin to transition into brown dwarf classification). Jupiters are one ideal class of exoplanets for our searches: they have large radii, and due to their lower mass compared to the next class, their internal heat flow can be very low. The minimum temperature expectation for Jupiters with masses and radii comparable to Jupiter, after 1 Gyr, is about $160$~K~\cite{caballero2018review}. After 10 Gyr, Jupiters are about $80$~K~\cite{caballero2018review}. Note that large cold gas giant planets are expected to be common~\cite{Gould:2006qb,Cassan_2012}. 
 \vspace{-4mm}
 
\subsubsection{Brown Dwarfs}
Larger again are brown dwarfs, which were only discovered in 1995. Brown dwarfs are what lurk in the gap between gas giant planets and the least massive stars, placing them with masses of about $14-75$ Jupiters. They generally have about the same radius as Jupiter, making them \textit{immensely dense}. This makes brown dwarfs an ideal candidate for our searches; they are large \textit{and} dense. However, for our scenarios of interest we do not want to solely consider very massive brown dwarfs; too massive and they take longer to cool. This leads to a large heat background that might obscure DM heating signals. The minimum temperature expectation for brown dwarfs with masses of about $14-75$, after 1 Gyr, is about $200-2000$~K respectively~\cite{Saumon_2008,Leggett_2017,caballero2018review}. After 10 Gyr, they range from about $150-1500$~K respectively~\cite{Saumon_2008,Leggett_2017,caballero2018review}. This means that while brown dwarfs of masses up to about 55 Jupiters can be relevant for our study, internal heat from the heavier dwarfs may outpower DM heating in some DM densities/locations, or would only be relevant for our study if their age approached 10 Gyr. It will therefore depend on the candidate in question, whether its individual heat background is acceptable or not. In any case, the abundance of cold brown dwarfs is expected to be very high~\cite{Cassan_2012,Reyl__2010}. More broadly, about 20$\%$ of stars are expected to have Jupiter-sized to brown dwarf sized planets~\cite{Cassan_2012}. Interestingly, brown dwarfs can have very exotic atmospheres, with some experiencing \textit{iron rain}~\cite{Buenzli_2012,Biller_2013}.

Brown dwarfs have been previously considered alongside DM, although in the context of asymmetric DM, which does not feature an annihilation heating signal, but rather a departure from the expected stellar evolution curve~\cite{Zentner:2011wx}. This is a similar approach to studying DM effects on stars in larger DM densities, such as in for example Refs.~\cite{1989ApJ346284B,1989ApJ33824S,Moskalenko_2007,Fairbairn:2007bn,Spolyar_2008,Freese_2008,Scott:2008ns,Taoso_2008,Freese:2008hb,Freese:2008ur,Iocco:2008xb,Freese_2010,Hooper:2010es,Casanellas_2010,Zackrisson_2010,Iocco:2012wk,Ilie_2012,Lopes:2014xaa,Lopes_2019,Hassani_2020}. However, these have different observables to those pointed out in this work.  \vspace{-4mm}

\subsubsection{Lost in Space: Rogue Planets}
Not all planets have a home. A class of planets called ``rogue planets'' or ``free-floating planets'' have been ejected from their planetary nursery, damned to aimlessly wander, alone, through dark and empty space. While all planet types listed above can be rogue planets, Jupiters and brown dwarfs are by far the most common rogues. This lonely class of exoplanets is ideal for our searches. This is because they are free from light and heat pollution from any host star, allowing them to be more easily resolved. Similarly, at closer distances to Earth, Jupiters on larger orbits can be easier to distinguish than those closely bound to their star, for the same reason.

While rogue planets are currently thought to be less common than bound planets, they can still be extremely plentiful. The OGLE survey estimates that there is up to about one rogue for every 4 stars -- that amounts to up to about 100 billion rogues in the galaxy~\cite{Mr_z_2017}. Even more recently, a simulation of planetary systems in the Orion Trapezium Cluster showed that about $15\%$ of all planets ended up ejected from orbit around their parent star~\cite{van_Elteren_2019}. (Interestingly, about $0.1\%$ ended up being later welcomed into a new family, captured by another distant parent star.) Extrapolating this system, it implies that there could be about 50 billion rogue planets in our Galaxy~\cite{van_Elteren_2019}. Alternatively, brown dwarfs can have never had a host star -- they can form in molecular clouds like stars, and simply be all alone from the very beginning.
 
\subsection{Candidates and Further Discovery Potential}
A number of promising exoplanet candidates have already been discovered in our galactic neighborhood, and extending to over 8 kpc away. In this section we summarize potential candidates that may be targets, as well as prospects for discovering new candidates in the future.

We emphasize that the observational exoplanet program is growing fast, and many more exoplanets will be characterized and discovered soon. Telescopes such as the James Webb Space Telescope (JWST), Transiting Exoplanets Survey Satellite (TESS), the Vera C. Rubin Observatory (Rubin), and the Nancy Grace Roman Space Telescope (Roman), and the Gaia Spacecraft have or will have targeted programs to discover as many exoplanets as possible. There are also many surveys such as the Optical Gravitational Lensing Experiment (OGLE), Two Micron All Sky Survey (2MASS), and the Wide-field Infrared Survey Explorer (WISE), which peer deep into our Galaxy. Further on the horizon, new telescopes are being planned or considered such as the Thirty Meter Telescope (TMT), the Extremely Large Telescope (ELT), Gaia Near Infra-Red (GaiaNIR), Large Ultraviolet Optical Infrared Surveyor (LUVOIR), Habitable Exoplanet Imaging Mission (HabEx), and the Origins Space Telescope (OST).

\subsubsection{Local Planets}

Table~\ref{tab:jovplanets} lists some known Jovian planets within 100 pc, which are potential candidates for the local exoplanet search. These are chosen as examples based on their proximity, radii, masses, and orbital sizes. JWST may be able to image these planets, and probe new DM parameter space. Note however that some may turn out to have too much atmospheric cloud cover, or may be heated or obscured for other reasons. Regardless, there are many more potential candidates, which can be found in Ref.~\cite{catalog}. 

In addition to known candidates, many current and future telescopes will study our local neighborhood to identify and measure more candidate planets for DM heating. In particular, Gaia is expected to find $21,000\pm6,000$ long-period Jupiters and brown dwarfs within 500 pc, within 5 years of operating~\cite{Perryman_2014}. Within 10 years, it is estimated to find $70,000\pm20,000$ new exoplanets of interest~\cite{Perryman_2014}. This will substantially increase candidates and statistics for this search.

\begin{table*}[]
\begin{tabular}{|cccccccc|} \hline
\hspace{1mm} Planet \hspace{1mm} &  Radius ($R_\textrm{jup}$)  &  Mass ($M_\textrm{jup}$) & \hspace{1mm} Distance \hspace{1mm} & \hspace{1mm} Orbit \hspace{1mm} & Temp (No DM) & Temp (with DM) &\hspace{1mm} Ref \hspace{1mm} \\ \hline
Epsilon Eridani b & 1.21 & 1.55 & 3 pc & 3.4 au & $\lesssim200$ K & $\lesssim650$ K &   \cite{Benedict_2006}\\
Epsilon Indi A b  & 1.17 & 3.25 & 3.7 pc & 11.6 au & $\lesssim200$ K & $\lesssim650$ K &   \cite{feng2018detection}\\
Gliese 832 b & 1.25 & 0.68 & 4.9 pc & 3.6 au & $\lesssim200$ K & $\lesssim650$ K &   \cite{Bailey_2008}\\
Gliese 849 b & 1.23 & 1.0 & 8.8 pc & 2.4 au & $\lesssim200$ K & $\lesssim650$ K &   \cite{Butler_2006}\\
Thestias & 1.19 & 2.3 & 10 pc & 1.6 au & $\lesssim200$ K & $\lesssim650$ K &   \cite{Hatzes_2006}\\
Lipperhey & 1.16 & 3.9 & 12.5 pc & 5.5 au & $\lesssim200$ K & $\lesssim650$ K &   \cite{Marcy_2002}\\
HD 147513 b & 1.22 & 1.21 & 12.8 pc & 1.3 au & $\lesssim200$ K & $\lesssim650$ K &   \cite{Mayor_2004}\\
Gamma Cephei b & 1.2 & 1.85 & 13.5 pc & 2.0 au & $\lesssim200$ K & $\lesssim650$ K &   \cite{Hatzes_2003}\\
Majriti & 1.16 & 4.1 & 13.5 pc & 2.5 au & $\sim218$ K & $\lesssim650$ K &   \cite{Butler_1999}\\
47 Ursae Majoris d  & 1.2 & 1.64 & 14 pc & 11.6 au & $\lesssim200$ K & $\lesssim650$ K &   \cite{Gregory_2010} \\ 
Taphao Thong  & 1.2 & 2.5 & 14 pc & 2.1 au & $\lesssim200$ K & $\lesssim650$ K &   \cite{Gregory_2010} \\
Gliese 777 b & 1.21 & 1.54 & 15.9 pc & 4.0 au & $\lesssim200$ K & $\lesssim650$ K &   \cite{Naef_2003}\\
Gliese 317 c & 1.21 & 1.54 & 15.0 pc & 25.0 au & $\lesssim200$ K & $\lesssim650$ K &   \cite{Johnson_2007}\\
q$^1$ Eridani b & 1.23 & 0.94 & 17.5 pc & 2.0 au & $\lesssim200$ K & $\lesssim650$ K &   \cite{Butler_2006}\\
HD 87883 b & 1.21 & 1.54 & 18.4 pc & 3.6 au & $\lesssim200$ K & $\lesssim650$ K &   \cite{Fischer_2009}\\
$\nu^2$ Canis Majoris c & 1.24 & 0.87 & 19.9 pc & 2.2 au & $\lesssim200$ K & $\lesssim650$ K &   \cite{Luque_2019}\\
Psi$^1$ Draconis B b & 1.21 & 1.53 & 22.0 pc & 4.4 au & $\lesssim200$ K & $\lesssim650$ K &   \cite{Endl_2016}\\
HD 70642 b & 1.19 & 1.99 & 29.4 pc & 3.3 au & $\lesssim200$ K & $\lesssim650$ K &   \cite{Carter_2003}\\
HD 29021 b &  1.2 &  2.4  & 31 pc & 2.3 au & $\lesssim200$ K & $\lesssim650$ K &   \cite{Rey_2017} \\ 
HD 117207 b &  1.2 &  1.9  & 32.5 pc & 4.1 au & $\lesssim200$ K & $\lesssim650$ K &   \cite{Marcy_2005} \\ 
Xolotlan & 1.2   &  0.9  & 34.0 pc & 1.7 au & $\lesssim200$ K & $\lesssim650$ K &   \cite{Vogt_2002} \\ 
HAT-P-11 c &  1.2 &  1.6  & 38.0 pc & 4.1 au & $\lesssim200$ K & $\lesssim650$ K &   \cite{Yee_2018} \\ 
HD 187123 c & 1.2  &  2.0  & 46.0  pc & 4.9 au & $\lesssim200$ K & $\lesssim650$ K &   \cite{Wright_2009} \\ 
HD 50499 b &  1.2 &  1.6  &  46.3 pc & 3.8 au & $\lesssim200$ K & $\lesssim650$ K &   \cite{Marcy_2005} \\ 
Pirx & 1.2   &  1.1  &  49.4 pc &  0.8 au & $\sim200$ K & $\lesssim650$ K &   \cite{Niedzielski_2009} \\ 
HD 27631 b & 1.2  & 1.5   & 50.3 pc & 3.2 au & $\lesssim200$ K & $\lesssim650$ K &   \cite{Marmier_2013} \\ 
HD 6718 b & 1.2  &  1.7  & 51.5  pc & 3.6 au & $\lesssim200$ K & $\lesssim650$ K &   \cite{Naef_2010} \\ 
HD 72659 b &  1.2 &  3.9  &  52.1 pc & 4.8 au & $\lesssim200$ K & $\lesssim650$ K &   \cite{Butler_2003} \\ 
HD 4732 c & 1.2  &  2.4  &  54.9 pc & 4.6 au & $\lesssim200$ K & $\lesssim650$ K &   \cite{Sato_2012} \\ 
HD 290327 b &  1.2 &  2.4  & 56.4  pc & 3.4 au & $\lesssim200$ K & $\lesssim650$ K &   \cite{Naef_2010} \\ 
HD 154857 c & 1.2  &  2.6  &  63.5 pc & 5.4 au & $\lesssim200$ K & $\lesssim650$ K &   \cite{Wittenmyer_2014} \\ 
Drukyul & 1.2  &  1.6  & 83.4  pc & 2.9 au & $\lesssim200$ K & $\lesssim650$ K &   \cite{Valenti_2009} \\ 
Kepler-539 c & 1.18 & 2.4  &  92 pc & 2.7 au & $\lesssim200$ K & $\lesssim650$ K &   \cite{Mancini_2016} \\ \hline
\end{tabular}
\caption{List of some candidate Jupiters within 100 pc, to use in a near-Earth search. Distance is quoted as from the Earth. The predicted temperature ranges include generic estimates for emissivity and planetary mass. Masses, radii, orbits, and distances from Earth are estimates taken from the NASA exoplanet catalog~\cite{catalog}.}
\label{tab:jovplanets}
\end{table*}

\vspace{-4mm}

\subsubsection{Distant Planets}
\vspace{-1mm}

The furthest planets ever found are SWEEPS-4 and SWEEPS-11~\cite{Sahu_2006}, which are about 8.5 kpc away (further than the Earth-GC distance). However, these planets are close to their host star, so are expected to be very hot (and therefore not ideal for DM heating searches). Many other planets are already known to exist, over varying distances from the GC. However, many of these planets are bound to a star.  While this is helpful for discovery techniques (i.e. more techniques are available to discover planets bound to a star), this is not helpful for our searches in the galactic bulge. This is because, even while they still may have very large orbits, at the very far distances into the GC that we want to measure, they can be outshone by their bound host star, making temperature measurements impossible. We therefore focus on rogue planets when examining potential DM signals at large distances.

While rogue planets are harder to find, some have already been found, and it is expected that many more can be found soon. Such searches require use of gravitational microlensing, which can be aided especially with simultaneous use of telescopes, allowing for more decisive confirmation of planetary status. For example, this has been achieved with Roman and Euclid~\cite{Bachelet_2019}.

OGLE has been operating since 1992, focusing on searches in the stellar bulge. It has already identified many distant exoplanets and exoplanet candidates. For rogue planet candidates, this includes e.g. OGLE-2019-BLG-0551~\cite{Mr_z_2020}, and a brown dwarf candidate OGLE-2015-BLG-1268, with 50 Jupiter masses and at $5.9\pm1.0$ kpc~\cite{Zhu__2016}. Spitzer has observed a candidate together with OGLE, called OGLE-2017-BLG-0896, which is potentially a brown dwarf about 4 kpc towards the bulge~\cite{Shvartzvald_2019}. OGLE is still very active, and will be important for identifying more candidates in the future.

In the near future, Roman is expected to find about 2200 new cold planets towards the galactic bulge (2000 of which have mass greater than Earth)~\cite{green2012widefield}, and hundreds of free-floating planets~\cite{Johnson_2020}. Our exoplanet searches benefit from large statistical samples, which Roman could provide. It aims to perform a deep near-infrared survey of the Galactic sky, and upon identifying candidates, can inform infrared telescopes such as JWST where to measure the temperature of the candidate planet. Alternatively, if a K-filter is added to Roman (allowing it to see further into the infrared) it itself may be able to measure the temperature of colder exoplanets~\cite{stauffer2018science}. 

\subsection{Temperature and Density Profiles}
To study the minimal DM mass that can be captured by the exoplanets we consider, a model of the conditions close to their core is required. To study Jupiters, we use the profile models for our Jupiter, as per Ref.~\cite{TilmanSpohn}. This features a core temperature of $T_c = 1.5\times10^4 \, \rm  K$, an average density of $\rho_{\rm jup} = 1.3 \,  \rm g/cm^3$, and a radius of $R_{\rm jup} = 6.99\times10^7 \,\rm m$. We set our benchmark Jupiters to all have the same radius as Jupiter. We also check if results vary with two different Jupiter density profile hypotheses, one with a core and the other without~\cite{TilmanSpohn}. For our parameters of interest, there is no noticeable effect.

To study brown dwarfs, we use the analytical model from Ref.~\cite{Auddy_2016}. The brown dwarf radius, core density and core temperature can be expressed as a function of mass and electron degeneracy, 
\begin{align}
\label{eqn:BD}
& R = 2.81 \times 10^9 \left( \frac{M\textsubscript{\(\odot\)}}{M} \right)^{1/3}\, \mu_e^{-5/3} \left( 1+ \gamma + \alpha \psi \right)\, \rm cm \, ,\\
 & \rho_c = 1.28 \times 10^5 \left(\frac{M}{M\textsubscript{\(\odot\)}}\right)^2\, \frac{\mu_e^{5} }{\left( 1+ \gamma + \alpha \psi \right)^3 }\,\, \rm g/cm^3 \, ,\\
 & T_c = 7.68 \times 10^8  \left( \frac{M}{M\textsubscript{\(\odot\)}} \right)^{4/3}\, \frac{\mu_e^{8/3} \psi }{\left( 1+ \gamma + \alpha \psi \right)^2 }\, \rm K \,. 
\end{align}
Here, $\mu_e$ is the number of electrons per baryon, $\psi$ the electron degeneracy parameter and $\gamma$ is a higher-order correction factor (see Ref.~\cite{Auddy_2016} for a detailed discussion). There is a particular electron degeneracy at which the core temperature reaches its maximum, and drops to smaller values if the degeneracy is further increased. Once the brown dwarf passes this point in the cooling process, its core temperature decreases significantly, while its density grows. The relatively low core temperatures and high densities make old brown dwarfs efficient accumulators for light DM. For our benchmark, the brown dwarf (BD) radius is taken to be $R_{\rm BD} = R_{\rm jup}$, and the mass $M_{\rm BD} = 55 \, M_{\rm jup} = 0.05 M_{\rm\odot}$.  This results in an average density of $\rho_{BD} = 73 \, \rm g/cm^3$, a core density of $\rho_c = 500 \, \rm g/cm^3$ and a core temperature $T_c = 2 \times 10^5 \, \rm K$. Note that the analytical model can give a range of brown dwarf core temperatures, depending on the brown dwarf properties and modeling assumptions. Also note that in the analytic model, fusion heating is not included. Our calculations are therefore only relevant in the regime where there is no fusion, which is appropriate to obtain lower internal heat backgrounds.

\section{Additional Search Details}
In this section we provide details relevant to our searches, in addition to the main text. This includes more details on a local search focusing on Super-Jupiters, and a search for distant exoplanets with a focus on rogue planets and brown dwarfs. We conclude with a subsection about possible challenges and point out opportunities for progress.

\subsection{Dark Matter Densities}

To calculate the DM-heating rate in exoplanets, we consider different DM profiles, which control the amount of DM available for heating at a given location in our Galaxy. We consider a Navarro-Frenk-White (NFW) profile, a generalized NFW (gNFW) profile, and a Burkert profile. The NFW profile is defined as a density as a function of galactic radius~\cite{Navarro_1996}
\begin{equation}
 \rho_\chi(r)=\frac{\rho_0}{(r/r_s)^\gamma(1+(r/r_s))^{3-\gamma}},
\end{equation}
where we take a scale radius of $r_s=8$~kpc~\cite{Lin:2019yux} (larger choices of e.g. $r_s=20$~kpc do not significantly change the results). The standard NFW profile has $\gamma=1$, while the generalized NFW profile is taken to have a steeper inner slope of $\gamma=1.5$ (this is equivalent to a Moore profile). This steeper value represents a more contracted profile, which can arise due to adiabatic contraction. Note that hydrodynamical simulations can sometimes produce even larger values of the inner slope~\cite{gnedin2011halo}.

Lastly, we also consider a cored profile, called the Burkert profile~\cite{Burkert_1995}, 
\begin{equation}
 \rho_\chi(r)=\frac{\rho_0}{(1+r/r_{\rm sb})(1+(r/r_{\rm sb})^2)}.
\end{equation}
For this profile, we will take a smaller core radius, such that $r_{\rm sb}=0.5$~kpc, to demonstrate reasonable variations in the profiles (as per Ref.~\cite{Cohen:2013ama}). Note however that in principle the core radius could be larger for this profile, making the DM density smaller towards the GC.

For all these profiles, we normalize to the local DM density value of $0.42$~GeV/cm$^3$~\cite{Pato_2015}. While we only consider variations of these profiles in our calculations, it is expected that overdensities -- localized regions of increased DM -- likely exist, and would be potentially detectable as hot exoplanets would deviate from the expectations of the profiles above, which we will also briefly investigate.

\subsection{Dark Heat Flow Compared with Background Internal Heat}

\begin{figure}[b]
\centering
\includegraphics[width=0.5\columnwidth]{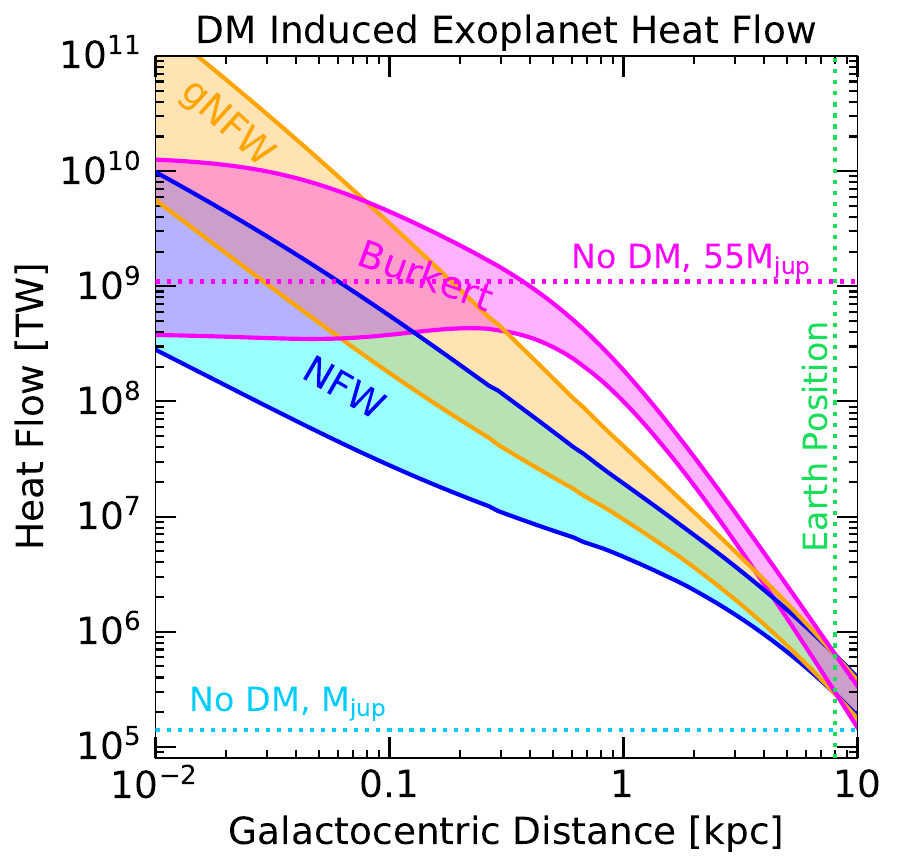}
\caption{Exoplanet heat power as a function of distance from the center of our Galaxy. Solid lines shown are the DM-heat power of exoplanets with radius $R_{\rm jup}$, with a range of masses from $M_{\rm jup}$ (lower line) up to $55\,M_{\rm jup}$ (upper line). The shaded region corresponds to DM-heat power for intermediate exoplanet masses. The dotted lines show the range of heat power for the heaviest dwarfs (with $55\,M_{\rm jup}$) down to the lighter benchmark Jupiters (with $M_{\rm jup}$) in the absence of DM or external heat.}
\label{fig:heat}
\end{figure}

Figure~\ref{fig:heat} shows the calculated heat flow from DM or internal heat, as a function of galactic radius. DM-heating arising due to several different DM profiles is shown, for NFW, gNFW, and Burkert profiles. The lower line shows the heat power prediction for exoplanets of mass $M_{\rm jup}$, while the upper line shows DM-heating for heavier brown dwarfs ($55\,M_{\rm jup}$). All planetary radii $R$ are taken to be $R_{\rm jup}$, which is the radius for all old brown dwarfs or Jupiters. Here, it is assumed that all of the DM passing through the planet is captured ($f = 1$). Any sub-maximal DM capture would lead to a heat power simply rescaled linearly with $f$. The shaded region for a given DM profile shows the intermediate temperatures for any Super Jupiter or lighter brown dwarf. The dotted lines show the internal heat for the two benchmark cases ($M_{\rm jup}$ vs $55\,M_{\rm jup}$) without DM heating, after 10 Gyr. Intermediate exoplanet masses without DM heating will fall between these lines. We see that for Jupiters (lower solid line), the DM heat will outperform the internal heat at all radii, making them ideal candidates for all searches. Brown dwarfs, on the other hand, being more dense, have higher internal heat, and DM heating will only clearly outperform their internal heat for some DM densities and radii.

The shape of the curves in Fig.~\ref{fig:heat} as a function of galactic radius is due to an interplay of the DM density profile, the DM velocity profile, and the effective capture radius of the exoplanet, which varies as a function of the DM velocity.

Note that another type of potential DM heating is kinetic DM heating. This arises when astrophysical systems have steep gravitational wells, causing DM to be accelerated to speeds near that of light. However, even for dense brown dwarfs, the escape velocity is only about $0.001c$, rendering any DM kinetic heating negligible.

\subsection{Local Search}

Nearby Jupiters, especially if DM heated, are within reach of direct imaging with JWST. This is an interesting possibility, as the gas giants of our own solar system are not well understood, and it could corroborate any potential DM contributions to their internal heat. Alternatively, measurement of some number of sufficiently cold exo-Jupiters would exclude this hypothesis, and allow for a DM scattering constraint to be set.

\begin{figure}[b]
\centering
\includegraphics[width=0.5\columnwidth]{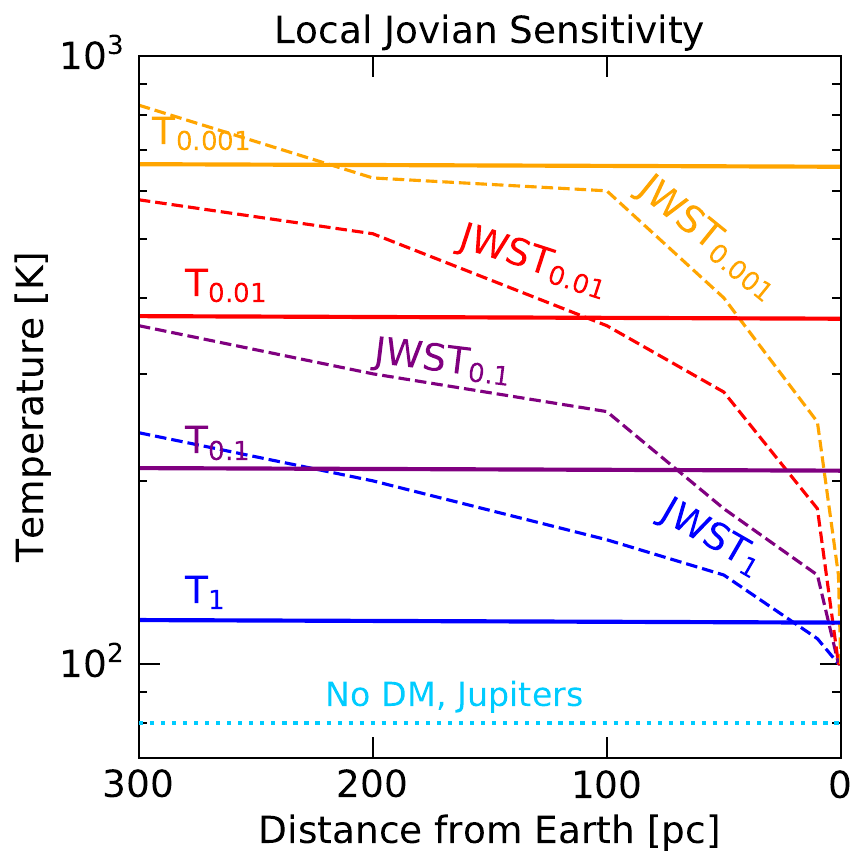}
\caption{Expected temperatures of local Jovian planets with masses of $M_{\rm jup}$. Increased temperatures due to DM-heating and varying emissivity values $\epsilon$ are shown as $T_\epsilon$ (solid lines), where distances shown are towards the GC. Corresponding best JWST sensitivities for each emissivity value are shown as JWST$_\epsilon$, as a function of the target-earth distance (dashed lines). We also show the temperature of an $\epsilon=1$ Jupiter without DM heating (dotted).}
\label{fig:jov}
\end{figure}

Figure~\ref{fig:jov} shows the expected exoplanet temperature as a function of distance from the center of the Galaxy, for Jupiters with different emissivities. We also show the optimal minimal JWST temperature sensitivity, for varying emissivity values. The JWST lines are found using the \textsc{MIRI: Imaging} instrument, with 2 SNR in $10^5$ seconds. Anything above this line requires comparable or shorter exposure times, for the given emissivity. For example, for a Jupiter within 10 pc, only about 50 seconds of exposure time is needed in the maximally DM-heated scenario (650 K, emissivity 0.001) to achieve 10 SNR (using \textsc{NIRISS} in \textsc{Imaging} mode). At 100 pc, 10 SNR can be achieved in about $10^5$ seconds. The weakened JWST sensitivities for decreasing emissivity values are due to the spectral flux having a lower normalization proportional to emissivity; see Appendix~\ref{app:emiss} for more details. The dotted line shows the temperature of a 10 Gyr Jupiter with no DM heating, for emissivity equal to one. The non-DM temperature for smaller emissivities may scale proportionally as $T/\epsilon^{1/4}$, however, the emissivity value may be affected by feedback effects in the cooling process. Note that even cold (non-DM heated) Jovian planets have been found to be detectable, in a more detailed JWST potential sensitivity analysis~\cite{Brande_2019}.

While the brown dwarf internal heating overpowers DM heating given local DM density, making them non-optimal targets for local searches, their high internal heat can be advantageous for another search strategy. The local relative abundance of dwarfs compared to other stellar populations, and the age/temperature distribution of the dwarfs can be determined, and can be potentially used to extrapolate to expected temperature/abundances towards the GC. This will be relevant for the distant search, as this would generate an apparent overabundance of younger brown dwarfs.

\subsection{Distant Search}

To demonstrate sensitivity to an example DM-heated large-distance exoplanet, we consider an exoplanet with radius $R_{\rm jup}$ at a distance from Earth of 8.2 kpc, which sits off the plane by 0.1 kpc. Given its location, for a mass of about $14\,M_{\rm jup}$, a DM-heated temperature of roughly 800~K would be obtained (assuming the gNFW profile), leading to a wavelength of $\nu^{-1}=3.6$ microns, and a flux density of $f_\nu=1.4$ nJy. This can be in principle measured by JWST using the \textsc{NIRISS: Imaging} mode, and the F356W filter, at 4 SNR with about $10^5$ seconds of exposure. With about $10^6$ seconds of exposure and the same filter, this can be detected at 10 SNR. In comparison, without DM heating, this exoplanet would have a temperature of around 220 K if sufficiently old and isolated, and so such a large increase in temperature would present as evidence for DM heating.

\begin{figure}[h!]
\centering
\includegraphics[width=0.5\columnwidth]{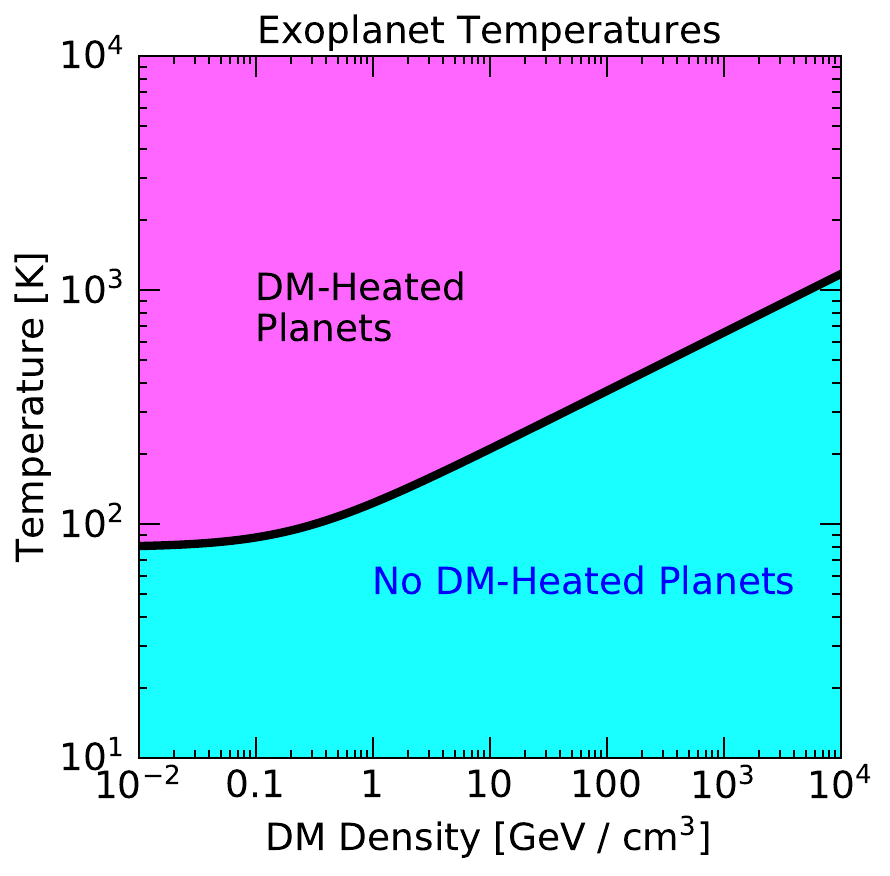}
\caption{Minimum DM-heated exoplanet temperature as a function of DM density, for a Jupiter-like planet with mass $M_{\rm jup}$ and radius $R_{\rm jup}$. A positive signal would lead to all planets being found in the ``DM-Heated Planets'' region, and none in the ``No DM-Heated Planets'' region. 
}
\label{fig:density}
\end{figure}

Figure~\ref{fig:density} demonstrates more generally how DM densities may be identified or excluded with exoplanets. For the Jupiter benchmark shown, no planets should be found in the DM densities given with the low temperatures as shown. On the other hand, all planets measured would be found in the DM-heating region. Any measurement contrary to this prediction would lead to an exclusion. In this figure, the DM velocity is set to $230\,$km/s -- the local DM velocity around a given candidate would need to be scaled in, along with gravitational focusing if the exoplanet is present in a low-DM velocity environment. The emissivity of the exoplanets in this plot it set to 1; this is a conservative choice, as lower emissivity values simply lead to higher DM-heating values. Note that overdensities of DM are expected from $N$-body simulations, such as \textsc{Via Lactea}~\cite{Diemand_2007,Diemand_2008}, \textsc{Ghalo}~\cite{Stadel_2009}, and \textsc{Aquarius}~\cite{Springel_2008}. It has been shown that the dense cores of many of the merging haloes that made our Milky Way survive as DM subhaloes, leading to e.g. DM streams or clumping. These overdensities or subhaloes can potentially be independently identified by gravitational microlensing (see e.g. Ref.~\cite{Metcalf:2001ap}).

\subsection{Challenges and Opportunities}

We have presented optimal estimates for JWST sensitivity, which take into account some effects that degrade the sensitivity. We now briefly discuss assumptions and estimates of the impact of these effects, and discuss directions to overcome some of these challenges.

\subsubsection{Exoplanet-Star Separation}

Bound exoplanets, at large distances, will invariably be too small to resolve with JWST, and will be outshone by their parent star. As such, we only consider free-floating planets for the distant search. However, for the local search, it is possible with JWST to observe bound exoplanets, as they can be close enough to be resolved. Regardless, planets on wider orbits are still preferable candidates even for the local search, as they are easier to disentangle from their star. An additional bonus for targeting planets on wider orbits, is that they are likely to have lower temperatures (as they receive less heat from their star).

A relevant aspect of the planet star separation is the question whether gravitational focusing of the parent star can affect the DM capture rate of the planet. We now briefly estimate the size of this effect. We assume as an estimate the exoplanet is orbiting a solar-like star at a distance of one to ten AU, similar to Jupiter. Exoplanets in much closer orbits would likely be heated significantly by the star and not be useful for our search.

The gravitational focusing effect can be written in terms of the parent star radius $R$, its escape velocity from the surface $v_{\rm esc}$ and the distance of the planet to the parent star $r$. The parameter controlling the strength of the effect is $ \epsilon = \frac{3}{2}\left(\frac{ v_{\rm esc}(r)}{ v_{\rm DM}} \right)^2 = \frac{3}{2}\left(\frac{ R\, v_{\rm esc}}{r \, v_{\rm DM}} \right)^2$, which increases the effective area as $A_{\rm eff} \approx R^2(1 + \, \epsilon)$. This means that even for a massive star with $v_{\rm esc} \sim 10^{-2}$ in natural units and $R \sim R_{\rm Sun}$, the correction factor is $\epsilon \approx 10^{-3} - 10^{-4}$ and thus negligible for planets in distant orbits. For comparison, the same parameter for the gravitational focusing of the exoplanet itself is in the range $\epsilon \approx 10 - 0.1$. 
The size of this estimate is approximately consistent with the numerical results in Refs.~\cite{Peter:2009mm,Peter:2009mi}, where the effects where studied in the solar system and have been found to amount to sub percent corrections at a distance of 1 AU from the Sun. This effect would become relevant for an exoplanet orbiting red giant star with a radius approaching sizes 100 times larger than the Sun, however for such a constellation, the planet would again be significantly heated by the star and not relevant for this search.

While we consider exoplanets on sufficiently wide orbits, assuming that the heat from the host star is negligible, note that this can be confirmed by measuring properties of the host star, which can be studied easily.

\subsubsection{Exoplanet Emissivities}

In Fig.~\ref{fig:jov}, we display a range of emissivities. These correspond to how efficiently the planet radiates heat; a lower emissivity will lead to more heat becoming trapped. While Fig.~\ref{fig:jov} showed fixed emissivities for each plot, in a true search, a given planet can have different emissivity values at different temperatures. To know where the planet would sit on an emissivity curve, it may be necessary to use complementary methods such as spectroscopy on the planet, to determine the atmospheric content, and therefore an estimate of the emissivity value.
For reference, Earth has an emissivity value of 0.6, our Jupiter is about 0.9, and Venus is about 0.004~\cite{Hooper:2011dw}.

In a similar vein, potential atmospheric clouds on an exoplanet can be important. Clouds may obscure the visibility of the planet, as was shown in simulations for JWST in Ref.~\cite{Brande_2019}. This size of this effect will depend on the candidate being observed, and can substantially vary from planet to planet.

\subsubsection{Determining Exoplanet Ages}

An independent age measurement of the exoplanet can certainly be a challenge in some circumstances. For brown dwarfs and Jupiters, while their temperature evolution curves are well known for a given mass and radius~\cite{Leggett_2017}, it may be difficult to differentiate between a hot younger exoplanet, and an older DM-heated exoplanet. The age of a given exoplanet can in principle be estimated from its surroundings. For example, if it is in a bound system, it can be calibrated from its parent star. Brown dwarfs can often be in a binary system with another type of star, which can allow the age of brown dwarfs to be estimated from their companion star~\cite{DES}.  If it is a free-floating exoplanet, calibration is substantially more difficult, but can in principle be deduced from the age of any nearby systems in which it may have originated. 

We expect overall that it will be difficult to resolve the age of all candidates, especially for planets at large distances. This could be due to, for example, no nearby system(s) that are easily enough identified as a rogue planet's ejecting host, or due to the age uncertainties simply being too large. We expect, however, that the large statistics that can be provided by these searches can form a sufficiently large dataset, which appears anomalous on average, given the expectations for the numerous systems studied. Conversely, and perhaps an easier task, is to find a number of sufficiently cold exoplanets, in contrast with the expectations of DM heating. This would allow a constraint on DM properties to be set.

Going forward, it will be important to investigate in more detail the impact of DM in stellar or planetary systems. Detailed simulations of temperature curves as a function of age for brown dwarfs or Jupiters, when DM heating is present and taking into account several planetary effects, will be required to make more precise statements than the estimates presented in this work. More broadly, we also emphasize that it may not be valid to calculate a planet's age based on its temperature alone, as is often done in the literature. This is because, as we have shown, DM heating can increase exoplanet temperatures, providing a departure from the standard age-cooling curves.

\subsubsection{How Far Away Can We See DM-Heated Exoplanets?}

The main source of signal degradation towards the GC is the number of stars present per pixel. This is because the number of stars increases dramatically, and these can overcrowd and outshine the exoplanet's heat signal, making it impossible to detect. We now estimate how far into the GC we can observe a DM-heated exoplanet by comparing the expected stars-per-pixel with our regions of interest. 

Comparing with known stellar densities~\cite{Valenti_2016}, and taking a line of sight about 1 degree above the GC, the stellar mass is about $2 \times 10^8 M_{\odot}$ per square degree. To convert this into a number of stars per square degree, we break the mass up into the fraction of expected mass in different types of stars. The stellar mass function is dominated by M-type stars contributing about $76 \%$ of the stellar population. We therefore expect about $5 - 7 \times 10^8$ stars per square degree. Now, considering JWST's NIRISS instrument, which provides our leading sensitivity towards the GC, we note that the field of view is 2.2 by 2.2 arcmin$^2$. The NIRISS instrument has a single 2048 by 2048 pixel detector array with 65 milliarcsec pixels. This means there would be about $5 - 7 \times 10^4$ stars in the field of view of JWST, and therefore that with NIRISS, we expect about 0.15 - 0.2 stars per pixel when observing about 1 degree above the GC. This means that, about $85\%$ of the time, an exoplanet candidate at this distance could potentially be observed without any stars contaminating its pixel. We therefore expect that about 1 degree off the plane, that is, out to about 0.1 kpc, is required to collect large statistics for observing this signal. In principle, it may be possible to push this sensitivity further, at the cost of sacrificing statistics due to even further increased stellar crowding. The absolute sensitivity will be limited by the extremely dense region around the central black hole, with a radius of about 30 pc~\cite{Moulin:2017cgb} -- at this point, any observations are likely completely hopeless. We however present only 0.1 kpc in our results, as a more realistic sensitivity cutoff that may collect enough statistics.

Another important source of signal degradation can be dust extinction. This occurs when more dust is present, as it may absorb the light emitted from the exoplanet, or any other background star. Dust extinction is most prevalent for shorter wavelengths, where the scatter with cosmic dust is more likely. 2MASS has studied extinction in the $K$-band (near infrared) within 10 degrees of the GC~\cite{10.1046/j.1365-8711.2003.06049.x}. The inner 0.5 degrees have very high dust extinction. For within $1-5$ degrees, the south has 60 percent less extinction than the north, making it a better target. The south also features Baade's window~\cite{Dutra_2002}, which due to its low dust extinction allowed the SWEEPS exoplanets to be discovered. Near 1 degree, the dust extinction value in the $K$-band  can be as low as $A_K=0.1$, which will not significantly affect a signal in infrared, particularly not for temperatures between about $500-800\,$K.

We therefore expect, considering both the stars-per-pixel and dust extinction, above about 1 degree off the plane, that is, out to about 0.1 kpc, provides us with our maximum optimal expected JWST sensitivity. Of course, depending on the location of the candidate exoplanet, and the properties of the planet itself, elements of this search will vary. We expect that exoplanet experimentalists will be able to determine much more accurate results than the estimates we have presented here. However, we find that leading factors such as dust extinction and the background stellar numbers do not appear to completely conceal the potential DM-heating, by aiming for candidates off the plane, out to least 0.1 kpc. This is why we truncate our optimal sensitivity estimates in all our figures at 0.1 kpc.

\subsubsection{Beyond JWST}

It is possible that other telescopes may prove more fruitful than JWST in the future. While the current design of Roman has a red cutoff at 2.0 microns, it has been argued that extending the wavelength sensitivity further into the IR, by adding an $K$-filter to Roman, could allow infrared imaging of distant free-floating exoplanets~\cite{stauffer2018science}, which due to its larger field of view and possible larger survey times, could potentially outperform JWST. It is also possible that Gaia Near Infra-Red (GaiaNIR), a proposed successor of Gaia in the near infrared, may improve this sensitivity, along with other potential future telescopes LUVOIR or OST.

\section{Dark Matter Equilibration}
\label{sec:equil}

DM capture and annihilation must reach equilibrium in order for the heating process to be maximally effective~\cite{Griest:1986yu}. In this subsection, we show equilibrium can be expected to be reached in our exoplanets of interest.  As our sensitivity extends into the sub-GeV regime, we consider both the standard $2 \rightarrow 2$ annihilation process, as well as $3 \rightarrow 2$ annihilation processes, which can be key for light DM. Note that in principle also DM decay can lead to an energy injection equilibrium, however the timescale required for equilibration is of the order of Gyr. Such short lifetimes are ruled out by indirect detection, see e.g. Ref.~\cite{Mitridate:2018iag}.

\subsection{$\chi\chi \rightarrow$ SM + SM Annihilation Processes}
For a DM candidate that annihilates via a $2 \rightarrow 2$ process to SM particles, the annihilation rate is given by the volume integral
\begin{equation}
 \Gamma_{\rm ann}^{2 \rightarrow 2} = \int dV n_{\chi}^2 \langle \sigma_{\rm ann} v_{\rm rel}\rangle, 
\end{equation}
where $n_\chi$ is the DM number density, and $\langle \sigma_{\rm ann} v_{\rm rel}\rangle$ is the thermal averaged cross section, with $\sigma_{\rm ann}$ the annihilation cross section, and $v_{\rm rel}$ the relative DM velocity. 

The equilibrium number of DM particles in the object is found from the solution of the differential equation
\begin{equation}
\dot{N}_\chi =  C_{\rm cap} -C_{\rm evap} N_\chi - C_{\rm ann}^{2 \rightarrow 2} \, N_\chi^2\,,
 \end{equation}
where $N_\chi$ is the DM number, $C_{\rm cap}$ is the capture rate, $C_{\rm evap}$ the evaporation rate, and the annihilation coefficient is given by
\begin{equation}
 C_{\rm ann}^{2 \rightarrow 2} = \langle \sigma_{\rm ann} v_{\rm rel}\rangle/V_{\rm eff}^{2 \rightarrow 2}.
\end{equation}
The annihilation volume is $V^{2 \rightarrow 2}_{\rm eff} = V_1^2/V_2$, with the volume for a given species $j$ being 
\begin{equation}
 V_j = 4\pi \int_{0}^{R} e^{- j m_{\chi} \phi(r)/T(r)} r^2 dr.
\end{equation}
Here, $R$ is the radius of the exoplanet, $T(r)$ is the planetary interior temperature as a function of radius, $\phi(r)$ is the gravitational potential, and $r$ is the radius of the volume within the exoplanet.
The equilibration time scale is then given by 
\begin{equation}
 \tau = (C_{\rm ann} C_{\rm cap})^{-1/2}\,.
\end{equation}
This can be converted into a lower bound on the annihilation cross section,
\begin{equation}
 \langle \sigma_{\rm ann} v_{\rm rel}\rangle \geq V^{2 \rightarrow 2}_{\rm eff}/(C_{\rm cap} \tau^2).
\end{equation}

To determine the minimum annihilation rates for exoplanets, we take the local DM density to be $\rho_{\rm DM} \approx 0.4 \text{ GeV}/\text{cm}^3$ and the velocity dispersion $v_d \approx 280 \rm \, m/s$, and for the GC we take $\rho_{\rm DM} \approx 10^3 \text{ GeV}/\text{cm}^3$ and the velocity dispersion $v_d \approx 50  \, \rm m/s$ as benchmark values. For Jupiters, we find that $\langle \sigma_{\rm ann} v_{\rm rel}\rangle \geq 10^{-30}\left(m_\chi/\text{GeV}\right)^{-1} \, \rm cm^3/s$ for local DM parameters and $\langle \sigma_{\rm ann} v_{\rm rel}\rangle \geq 10^{-34}\left(m_\chi/\text{GeV}\right)^{-1} \, \rm cm^3/s$ for GC values, which is expected to be satisfied in models with a thermal freezeout. For a brown dwarf at the GC, the minimum annihilation rate is $\langle \sigma_{\rm ann} v_{\rm rel}\rangle \geq 10^{-37}\left(m_\chi/\text{GeV}\right)^{-1} \, \rm cm^3/s$.

\subsection{$\chi+\chi+\chi \rightarrow \chi+\chi$ Annihilation Processes}

As our searches focus on the sub-GeV regime, now we discuss models of light, thermally produced DM, which are based on $3 \rightarrow 2$ interactions. In a broad class of models, suggested in Ref.~\cite{Hochberg:2014dra}, the DM freezes out via the interaction $\chi + \chi + \chi \rightarrow \chi + \chi$. The freezeout condition for the interaction rate factor is then $\langle \sigma_{3 \rightarrow 2} v_{\rm rel}^2\rangle = 10^8 \left(m_\chi/\text{GeV}\right)^{-1} \rm GeV^{-5}$~\cite{Dondi:2019olm}. Thus this is the value of the interaction strength predicted from the thermal freezeout condition for a SIMP.  Note that the rate factor $\langle \sigma_{3 \rightarrow 2} v_{\rm rel}^2\rangle$ has the dimension of an area times a volume, such that when it is multiplied by the number density squared of incoming particles, the result is an interaction rate. The corresponding annihilation rate in planets is given by the volume integral 
\begin{equation}
 \Gamma_{\rm ann}^{3 \rightarrow 2} = \int dV n_{\chi}^3 \langle \sigma_{3 \rightarrow 2} v_{\rm rel}^2\rangle,
\end{equation}
resulting in an annihilation rate of
\begin{equation}
 C_{\rm ann}^{3 \rightarrow 2} =\langle \sigma_{3 \rightarrow 2} v_{\rm rel}^2\rangle/(V_{\rm eff}^{3 \rightarrow 2})^2,
\end{equation}
with an annihilation volume of $V^{3 \rightarrow 2}_{\rm eff} = V_1 \sqrt{V_1/V_3}$. The equilibrium condition on the rate factor then reads
\begin{equation}
 \langle \sigma_{3 \rightarrow 2} v_{\rm rel}^2\rangle \geq \left(V^{3 \rightarrow 2}_{\rm eff}\right)^2/(C_{\rm cap} \tau^2).
\end{equation}
Given a Jupiter-like planet with $M = M_{\rm jup}$, $R=R_{\rm jup}$ and $\tau = 10 \, \rm Gyr$, this gives 
\begin{equation}
 \langle \sigma_{3 \rightarrow 2} v_{\rm rel}^2\rangle \geq 10^{42} \left(m_\chi/\text{GeV}\right)^{-2}\, \rm GeV^{-5}
\end{equation}
at the local DM density and velocity and 
\begin{equation}
 \langle \sigma_{3 \rightarrow 2} v_{\rm rel}^2\rangle \geq 10^{37} \left(m_\chi/\text{GeV}\right)^{-2}\, \rm GeV^{-5}
\end{equation}
at the GC DM density and velocity. For a brown dwarf at the GC, the minimum annihilation rate is  $\langle \sigma_{3 \rightarrow 2} v_{\rm rel}^2\rangle \geq 10^{26} \left(m_\chi/\text{GeV}\right)^{-2}\, \rm GeV^{-5}$. This is many orders of magnitude larger than the value expected from the thermal freezeout, referenced above. This means that the $\chi+\chi+\chi \rightarrow \chi+\chi$ process does not reach equilibrium in sufficient time. There is however, another $3\rightarrow2$ process that can be relevant, which we now discuss. 

\subsection{$\chi+\chi+SM \rightarrow \chi+SM$ Annihilation Processes}

Recently, a different number changing interaction has been proposed in order to produce light, thermal DM, called the Co-SIMP~\cite{Smirnov:2020zwf}. In this scenario, the DM freeze-out is assisted by SM particles, in the process $\chi + \chi + \text{SM} \rightarrow \chi + \rm SM$. Since the number density of SM particles in a planet is by many orders of magnitude larger than the accumulated DM number density, this interaction rate is significantly more efficient. This leads to a prediction for the rate factor, $\langle \sigma_{3 \rightarrow 2} v_{\rm rel}^2\rangle = 10^3 \left(m_\chi/\text{GeV}\right)^{-3} \rm GeV^{-5}$. This is the value of the interaction strength predicted from the thermal freezeout condition for a Co-SIMP.  The annihilation rate in exoplanets is given by 
\begin{equation}
 \Gamma_{\rm ann}^{3 \rightarrow 2} = \int dV n_{\chi}^2 n_{\rm SM} \langle \sigma_{3 \rightarrow 2} v_{\rm rel}^2\rangle,
\end{equation}
resulting in an annihilation rate of
\begin{equation}
 C_{\rm ann}^{3 \rightarrow 2} = \langle \sigma_{3 \rightarrow 2} v_{\rm rel}^2\rangle  \, n_{\rm SM}/ V_{\rm eff}^{2 \rightarrow 2},
\end{equation}
with the condition that 
\begin{equation}
 \langle \sigma_{3 \rightarrow 2} v_{\rm rel}^2\rangle \geq V_{\rm eff}^{2 \rightarrow 2} /(\tau^2 \,C_{\rm cap} n_{\rm SM}).
\end{equation}
For a Jupiter-like planet, this gives a minimum rate to reach equilibrium,   
\begin{equation}
 \langle \sigma_{3 \rightarrow 2} v_{\rm rel}^2\rangle \geq  10^{-14} \left(m_\chi/\text{GeV}\right)^{-1}\, \rm GeV^{-5}
\end{equation}
at the local DM density and velocity and
\begin{equation}
 \langle \sigma_{3 \rightarrow 2} v_{\rm rel}^2\rangle \geq  10^{-18} \left(m_\chi/\text{GeV}\right)^{-1}\, \rm GeV^{-5}
\end{equation}
at the GC DM density and velocity. For a brown dwarf at the GC the minimum annihilation rate is $\langle \sigma_{3 \rightarrow 2} v_{\rm rel}^2\rangle \geq  10^{-24} \left(m_\chi/\text{GeV}\right)^{-1}\, \rm GeV^{-5}$. Those values are well below the thermally expected rate, such that captured Co-SIMP particles always reach equilibrium. As we consider elastic cross sections that lead to all the outgoing particles becoming trapped in the planet, this means that the entire mass energy released in the Co-SIMP process will be converted to the planetary heat flow. 

In fact, this process is even more broadly applicable. In Ref.~\cite{Hochberg:2014dra}, it is shown that in the SIMP model the Co-SIMP process exists, however, the rate is suppressed by a factor $\epsilon^2$. Experimentally this quantity is constrained to be in the range of $\epsilon \sim 10^{-6} - 10^{-8}$, and therefore the Co-SIMP process will be subdominant to the thermal SIMP rate. Regardless, the subdominant Co-SIMP process will still be the process that brings the particles into annihilation equilibrium (as opposed to kinetic equilibrium that only equilibrates the temperatures), due to the larger number density. We therefore expect that the Jupiter-like and brown dwarf searches will probe new territory of the number changing, thermal DM models.   

\section{Dark Matter Evaporation}
\label{sec:evap}

DM evaporation from the exoplanet truncates the low DM mass sensitivity. This is because in collisions with SM particles, the DM velocity of individual particles can exceed the escape velocity, and the particles evaporate from the object. We emphasize that the value of the evaporation mass strongly depends on the considered model. For concreteness, we give now some model-specific examples. We use the evaporation rate derived in~Ref.~\cite{Gould:1989tu}, given by
\begin{align}
\Gamma_{\rm evap.} = \frac{8}{\pi^3} \frac{\Sigma_{\rm total}}{\bar{r}^3} \bar{v} \frac{E_0}{T(\bar{r})} \, \exp{\left( - \frac{E_0}{T(\bar{r})} \right)},
\end{align}
where $\Sigma_{\rm total}$ is the sum of all the cross sections of scattering protons, $\bar{v}$ the thermal DM velocity, $E_0$ the DM escape energy, and $\bar{r}= \left( \frac{6 T(\bar{r})}{\pi^2 G_N \rho( \bar{r}) m_\chi} \right)^{1/2}$ is the mean DM radius. We evaluate the evaporation mass now in four different scenarios: the WIMP s-wave and p-wave annihilation, the Co-SIMP case, and a light mediator case, for a Jupiter-like planet and a $M_{\rm BD} = 55 \, M_{\rm jup} $ brown dwarf. We consider two benchmark cross sections, $10^{-34}$~cm$^2$ and $10^{-30}$~cm$^2$. We define the evaporation mass as the DM mass value where the evaporation time-scale is roughly shorter than the time-scale in which  annihilation equilibrium is reached, as per Sec.~\ref{sec:equil}. More precisely, the evaporation condition of Ref. \cite{Garani:2017jcj} is adopted i.e. $\Gamma_{\rm evap} \tau_{\rm eq} = 1/\sqrt{0.11}$ when annihilation equilibrium is reached, which singles out the rate at which DM depletion due to evaporation is at the $90\%$ value of the captured particle number.

To obtain evaporation estimates, we make some assumptions about the planetary properties. For Jupiter, we take a profile with a core temperature of $T_c = 1.5 \times 10^4 \, \rm  K$, an average density of $\rho_{\rm jup} = 1.3 \,  \rm g/cm^3$, and a radius of $R_{\rm jup} = 6.99\times10^9 \,\rm cm$. For our brown dwarf benchmark point we use an analytical model as detailed in Sec.~\ref{sec:overview}. Compared to our cross section sensitivity estimates, where we assumed a hydrogen sphere, the evaporation limits require a model for the exoplanet core. The relatively low core temperatures and high densities make old brown dwarfs efficient accumulators for light DM. For our benchmark, the brown dwarf radius is taken to be $R_{\rm BD} = R_{\rm jup}$, and the mass $M_{\rm BD} = 55 \, M_{\rm jup} = 0.05 M_{\rm\odot}$.  This results in an average density of $\rho_{BD} =  73 \, \rm g/cm^3$, a core density of $\rho_c = 500 \, \rm g/cm^3$ and a core temperature $T_c = 2 \times 10^5 \, \rm K$. Note that the analytical model can give a range of brown dwarf core temperatures, depending on the brown dwarf properties and modeling assumptions; it would be reasonable to also take BD core temperatures a factor few higher, which would result in evaporation masses a factor few higher (see e.g. Ref.~\cite{Acevedo:2023owd}). We do not study brown dwarfs higher than $M_{\rm BD} = 55 \, M_{\rm jup}$, as the internal background heat eventually becomes too high to see a DM heating signal. Any other Jupiters, Super Jupiters, or lighter brown dwarfs, will have a minimum exoplanet-bounded DM mass that is between the Jupiter benchmark (planetary mass $M_{\rm jup}$), and the brown dwarf benchmark (planetary mass $55\,M_{\rm jup}$). The DM mass where evaporation occurs will depend on where the mass of the exoplanet in question sits relevant to these benchmarks.

 Table~\ref{tab:evaporationmass} shows a range of evaporation masses in the context of some example DM models. We see that the value of the evaporation mass is strongly model dependent. This occurs because different models have different annihilation rates, and a faster annihilation rate can lead to lower retained DM masses, or because in some models both the DM distribution and the DM escape energy can change. The annihilation rates for the WIMPs are assumed as the standard freezeout values obtained from an s-wave or p-wave annihilation process. The Co-SIMP annihilation rate has no scaling with velocity, and the rate factor is determined by the freezeout value quoted above. Note that the thermal annihilation rate for Co-SIMPs can be faster than for WIMPs at high cross sections, leading to potential lower evaporation masses. For the light mediator case, $g_\chi$ corresponds to the DM-mediator coupling, $g_{\rm SM}$ is the SM-mediator coupling, and $m_\phi$ is the mediator mass. The condition quoted in the table, $g_\chi\,g_{\rm SM}\gtrsim 10^{-9}$eV$^2/m_\phi^2$, leads to evaporation masses as listed, under the assumption that the light mediator provides an evaporation barrier but does not provide the scattering rate; see Ref.~\cite{Acevedo:2023owd} for detailed discussion.\newpage
 
There are further aspects that can significantly affect the value of the evaporation mass, including:
\begin{itemize}
\item Larger elastic cross sections, where the DM is more efficiently trapped. This regime ideally requires simulations.
\item Self-interactions of DM, leading to additional DM trapping.
\end{itemize}
Further considerations are beyond the scope of our work, but leave plenty of room for investigations in specific particle DM models.

In a previous version of this manuscript, we took simple estimates of the evaporation mass for our plots using the condition that the average value of the DM velocity is less than the escape velocity. As argued in Ref.~\cite{Garani:2021feo}, this does not include the tail of the velocity distribution. However, neither our previous simple treatment, nor Ref.~\cite{Garani:2021feo}, consider model dependent effects. As was already emphasized in our previous manuscript version, the evaporation mass is highly-model dependent, and so we only included the simple estimate based on the average value of the DM velocity being less than the escape velocity in the plots of our previous manuscript version. To remove any confusion about the interpretation of our plots, we have removed any evaporation value from the plots.
Tab.~\ref{tab:evaporationmass} should be referenced for some examples of specific model evaporation values.

\begin{table*}[]
\begin{tabular}{|ccccc|} \hline
\hspace{1mm} Model \hspace{3mm} & \begin{tabular}{@{}c@{}}Jupiter\vspace{-2mm}\\ $\sigma = 10^{-34} \text{cm}^2$\end{tabular}\hspace{3mm} & \begin{tabular}{@{}c@{}}Jupiter\vspace{-2mm}\\ \hspace{1.5mm}$\sigma = 10^{-30} \text{cm}^2$\end{tabular}\hspace{1.5mm}  & \begin{tabular}{@{}c@{}}Brown Dwarf\vspace{-2mm}\\ \hspace{1.5mm}$\sigma = 10^{-34} \text{cm}^2$\end{tabular}\hspace{1.5mm} & \hspace{3mm} \begin{tabular}{@{}c@{}}Brown Dwarf\vspace{-2mm}\\ \hspace{1.5mm}$\sigma = 10^{-30} \text{cm}^2$\end{tabular}\hspace{1.5mm}  \\ \hline
Co-SIMP & $\sim 1.2 \text{ GeV}$ & $\sim 220 \text{  MeV} $ & $\sim 290 \text{  MeV}$  & $\sim 25 \text{  MeV}$ \\
WIMP s-wave & $\sim 1.1 \text{ GeV}$ &  $\sim 180 \text{ MeV}$  & $\sim 260 \text{  MeV}$  &  $\sim 30 \text{  MeV}$ \\
WIMP p-wave & $\sim 1.2 \text{ GeV}$ & $\sim 250 \text{  MeV}$ & $\sim 400 \text{  MeV} $ &  $\sim 40 \text{  MeV} $\\
\begin{tabular}{@{}c@{}}Light Mediator\vspace{-2mm}\\ $g_\chi\,g_{\rm SM}\gtrsim 10^{-9}$eV$^2/m_\phi^2$ \end{tabular} & $\lesssim 100 \text{ keV}$ & $\lesssim 100 \text{ keV}$ & $\lesssim 100 \text{ keV}$ &  $\lesssim 100 \text{ keV}$\\
 \hline
\end{tabular}
\caption{DM evaporation masses for different objects at different cross sections, in the local Galactic position, for several DM benchmark models. Other model variations can provide lower DM evaporation masses (see text).}
\label{tab:evaporationmass}
\end{table*}

\section{Additional Cross Section Sensitivity Results}
\label{sec:addcross}
In this section, we consider additional cross section sensitivities. We first provide additional details for the spin-dependent proton scattering results, including additional results in more DM velocity regimes, as well as the spin-independent results. We then show results for DM scattering with electrons in exoplanets, followed by some additional discussion of all the cross section results.

\subsection{Spin-Dependent DM-Proton Scattering Results in Local Velocities}
We now briefly provide additional details for the spin-dependent results. For these interactions, we parametrize the cross section in the following way:
\begin{align}
\sigma_{\chi A}^{\rm SD} =  \sigma_{\chi N}^{\rm SD} \left( \frac{\mu(m_A)}{\mu(m_N)} \right)^2 \frac{4(J+1)}{3 J} \left[ a_p \langle  S_p \rangle + a_n \langle  S_n \rangle \right]^2 \,,
\end{align}
where $J$ is the total nuclear spin, $ \langle  S_{p} \rangle$ and $ \langle  S_{n} \rangle$ the effective proton and neutron spins of the nucleus respectively, and $a_{p}$ and $a_{n}$ the model dependent DM-proton and DM-neutron coupling strength respectively. For our scenario only the couplings to protons will be relevant, since our targets are dominantly made of hydrogen and helium, and the latter has zero total nuclear spin. We assume $a_p = 1$. For simplicity, as per the SI limits, we approximate our targets as proton spheres.

\begin{figure*}[t!]
\centering
\includegraphics[width=0.45\columnwidth]{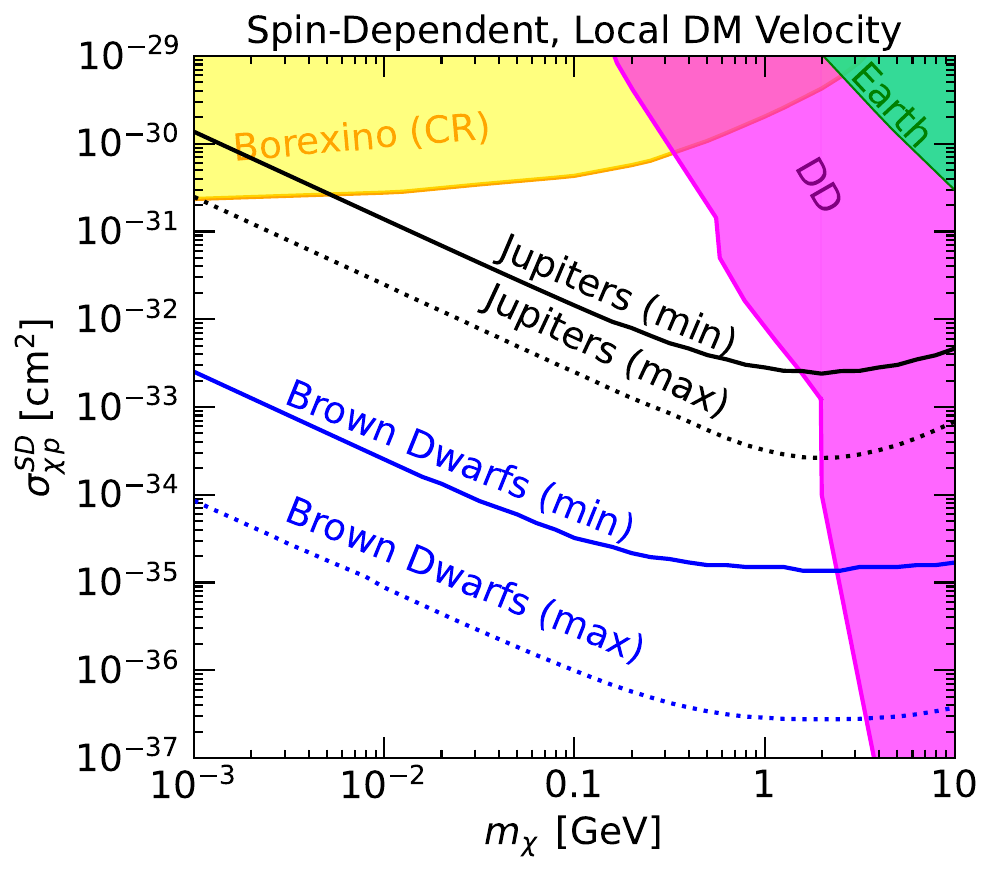}
\caption{Spin-dependent DM-proton scattering cross section sensitivity estimates for Jupiters and brown dwarfs, for exoplanets in a local DM velocity calculated in this work. The solid (min) lines show cross sections assuming effectively all DM is captured, and the dotted lines show the maximum expected reach. Complementary constraints are also shown; Earth is the limit on Earth DM-heat flow~\cite{Bramante:2019fhi}, DD is a collection of direct detection experiments~\cite{Agnese:2017jvy,Collar:2018ydf,Abdelhameed:2019hmk,Hooper:2018bfw,Aprile:2019jmx,Liu:2019kzq}, Borexino (CR)~\cite{Bringmann:2018cvk} corresponds to cosmic-ray boosted DM signals.}
\label{fig:spindeploc}
\end{figure*}

Figure~\ref{fig:spindeploc} shows our sensitivities to spin-dependent DM-proton scattering in local DM velocities. In the main text, we only showed the results for spin-dependent DM-proton scattering in GC DM velocities. We see that the scattering limits in the local DM velocities are not as strong; this is because the velocities are higher, and therefore there is much less boost from gravitational focusing. As per the main text results, the ``min'' cross section corresponds to the case where effectively all DM is captured (about $95\%$). The ``max'' cross section corresponds to the smallest DM capture fraction (about $10\%$) that can be probed in the near future with JWST. This is likely the maximum cross section reach, as the temperatures corresponding to this lower scattering rate are approaching either the JWST minimum temperature detection threshold, or the expected background temperature, in most of the parameter space. The lower end of the curves is truncated by the mass in which DM evaporates out of the system. Here, we have not included evaporation as this value is highly model dependent; see Sec.~\ref{sec:evap} for discussion.

Note that the scattering sensitivity arises predominately from DM-proton interactions. This is because gas giants and brown dwarfs are predominately hydrogen and helium; hydrogen only has a proton, and helium has zero total nuclear spin, thus DM-neutron interactions are not significant. We also show the earth heat flow bounds from Ref.~\cite{Bramante:2019fhi} for comparison, and direct detection bounds~\cite{Gangi:2019zib, Aprile:2019dbj, Aprile:2019jmx,Liu:2019kzq}. We also show limits on boosted DM from collisions with cosmic rays~\cite{Cappiello:2018hsu, Bringmann:2018cvk,Ema:2018bih, Cappiello:2019qsw}, shown as ``Borexino (CR)''. Note that these limits have different assumptions; the direct detection limits do not require any minimum annihilation cross section.

\subsection{Spin-Independent DM-Nucleon Scattering Results}

In the main text, we only showed the spin-dependent results. Here, we now show our spin-independent results.

\begin{figure*}[t]
\centering
\includegraphics[width=0.45\columnwidth]{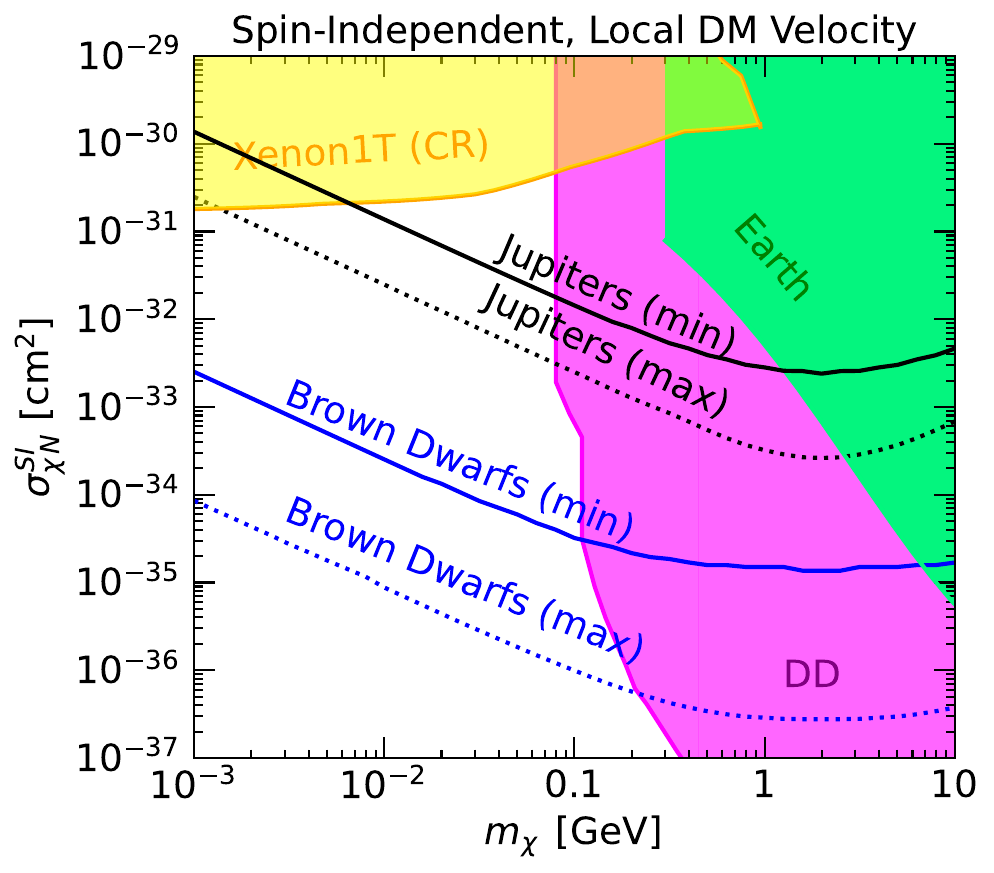}
\includegraphics[width=0.45\columnwidth]{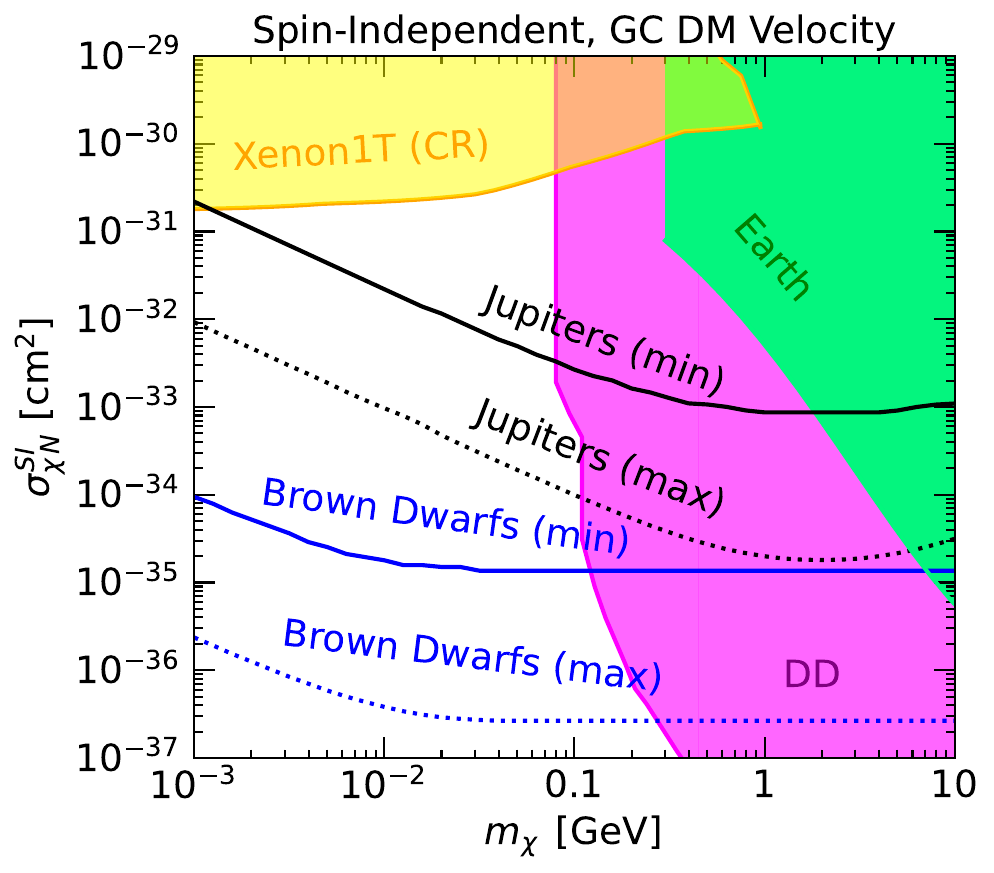}
\caption{Spin-independent DM-nucleon scattering cross section sensitivity estimates for Jupiters and brown dwarfs, for exoplanets in a local DM velocity (left column) or GC DM velocity (right column) calculated in this work. The solid (min) lines show cross sections assuming effectively all DM is captured, and the dotted (max) lines show the maximum expected reach. Complementary constraints are also shown; Earth is the limit on Earth DM-heat flow~\cite{Bramante:2019fhi}, DD is a collection of direct detection experiments~\cite{Agnese:2017jvy,Collar:2018ydf,Abdelhameed:2019hmk,Hooper:2018bfw,Aprile:2019jmx,Liu:2019kzq}, Xenon1T (CR)~\cite{Bringmann:2018cvk} corresponds to cosmic-ray boosted DM signals.}
\label{fig:spinindep}
\end{figure*}

Figure~\ref{fig:spinindep} shows our spin-independent results, for exoplanets in local and GC velocities. Compared to the spin-dependent results, the exoplanet sensitivities are the same, however the complementary constraints vary. Most notably, the direct detection constraints are stronger. The lower end of the curves is truncated by the mass in which DM evaporates out of the system. Here, we have not included evaporation as this value is highly model dependent; see Sec.~\ref{sec:evap} for discussion.

\subsection{DM-Electron Scattering Sensitivity}

To obtain the limits on DM-electron scattering in exoplanets, we also assume a hydrogen sphere for the exoplanets. As the chemical composition is dominantly hydrogen, this allows the assumption that the proton number density is identical to the electron number density.  A subdominant correction comes from the helium abundance, which we neglect to be conservative. Note that given the hydrogen target, relativistic shell effects play no role in the considered processes.  We assume a momentum independent DM-electron cross section $\sigma_{\chi \rm e}$, i.e. the electron form factor is $F=1$. 

\begin{figure*}[t!]
\centering
\includegraphics[width=0.45\columnwidth]{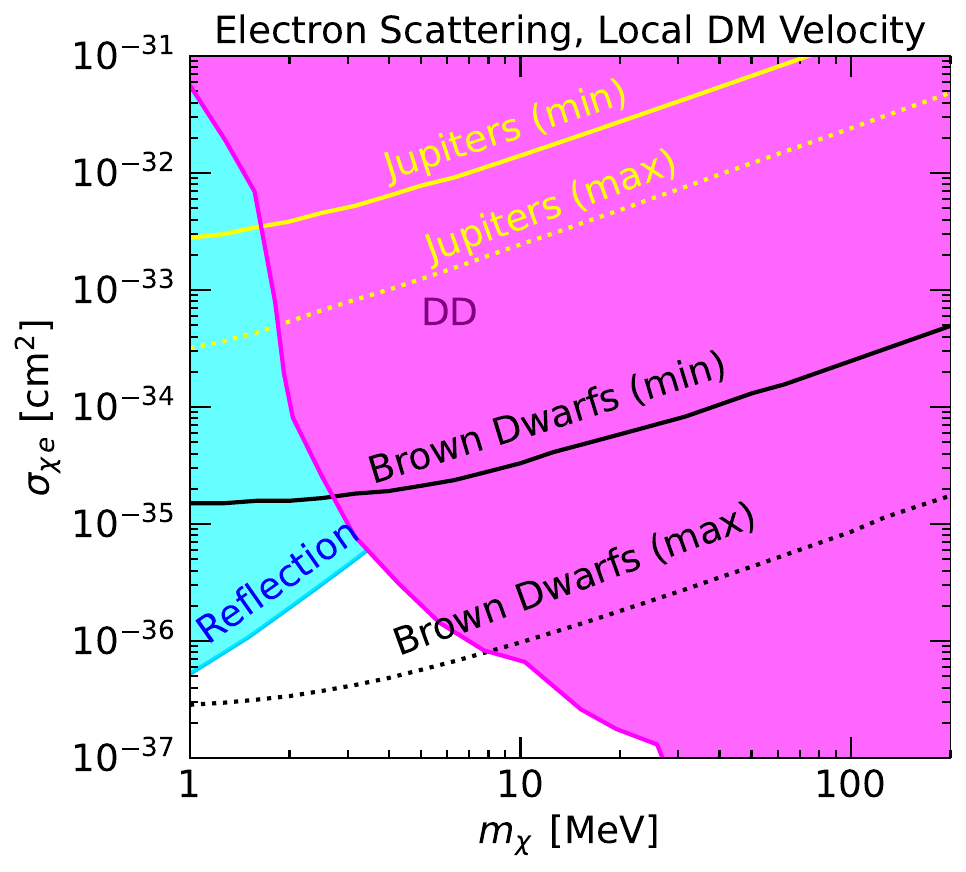}
\includegraphics[width=0.45\columnwidth]{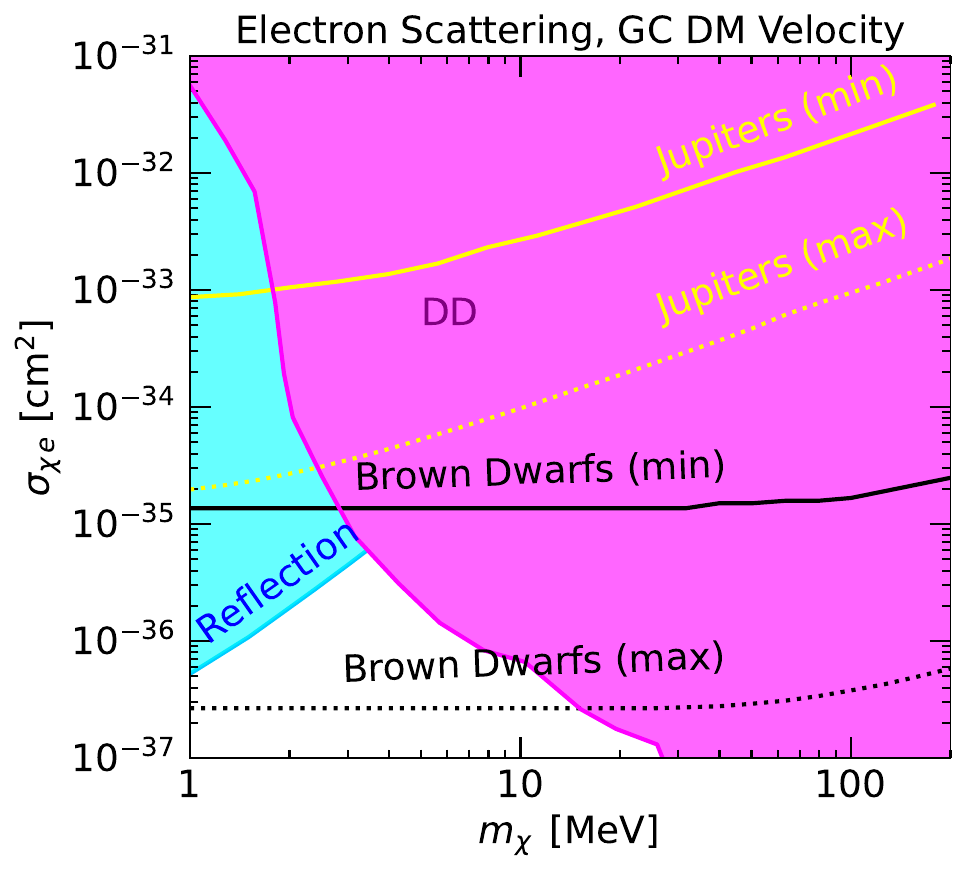}
\caption{DM-electron scattering cross section sensitivity for brown dwarfs and Jupiters, for exoplanets in a local DM velocity (left) or GC DM velocity (right) calculated in this work. The solid (min) lines show cross sections assuming effectively all DM is captured, and the dotted (max) lines show the maximum expected reach. Complementary constraints from direct detection (DD)~\cite{Essig_2012,Essig_2017,Angle_2011,Aprile_2016,Agnes_2018,Agnese_2018,Crisler_2018,Abramoff_2019,Aguilar_Arevalo_2019,barak2020sensei,collaboration2020germaniumbased,Amaral:2020ryn} and solar reflection~\cite{An:2017ojc} are shown.}
\label{fig:xseclep}
\end{figure*}

Figure~\ref{fig:xseclep} shows the DM-electron scattering sensitivity estimates, alongside existing limits from direct detection~\cite{Essig_2012,Essig_2017,Angle_2011,Aprile_2016,Agnes_2018,Agnese_2018,Crisler_2018,Abramoff_2019,Aguilar_Arevalo_2019,barak2020sensei,collaboration2020germaniumbased,Amaral:2020ryn} and solar reflection~\cite{An:2017ojc,Emken_2018}. Note that these limits have different assumptions; the direct detection limits do not require any minimum annihilation cross section. We show the sensitivity for when about all ($95\%$) and the likely smallest possible detectable amount ($10\%$) of DM is captured.  Electron-dominated interactions may be found in for example leptophilic DM models~\cite{Fox:2008kb,Kopp:2009et,Kopp:2014tsa,Bell:2014tta, DEramo:2017zqw}. The lower end of the curves is truncated by the mass in which DM evaporates out of the system. Here, we have not included evaporation as this value is highly model dependent; see Sec.~\ref{sec:evap} for discussion.

Note that the difference in results between scattering off different targets such as nuclei and electrons is simply due to their different masses. When the DM mass is comparable to the target mass, scattering is most efficient. This is why the shape of the electron scattering cross section plots and the nucleon scattering plots are different -- the nucleon (proton) scattering is most efficient around the proton mass of 1 GeV, and therefore strongest around this mass. On the other hand, the electron scattering is most efficient around the electron mass of about 0.5 MeV, and therefore the cross section limits are strongest approaching this mass.

\subsection{Additional Cross Section Results Discussion}

For all of Fig.~\ref{fig:spindeploc}, Fig.~\ref{fig:spinindep}, and Fig.~\ref{fig:xseclep} (and Fig.~3 of the main text), depending on the DM model, the sensitivity region could be all filled in as a constraint if a statistically significant number of sufficiently cold Jupiters or brown dwarfs were measured. For instead discovery of a DM-heating signal, the DM parameters would lie above the lines shown. As both these figures show $95\%$ (max) and $10\%$ (min) values of the DM capture rate, in principle even stronger sensitivity to DM cross sections can be reached if a smaller DM capture fraction can be probed. However, given the JWST optimal sensitivity, about a $10\%$ DM capture fraction is likely the smallest capture fraction that can be probed in the near future. Note that the choice of $95\%$ for the max is simply for definitiveness; a $95\%$ and $100\%$ capture rate heating signal cannot be discerned by the telescope, but the $100\%$ capture rate corresponds to an infinite cross section, which is not well-defined.

Note that there is a ceiling for the cross sections above which the DM does not drift fast enough into the planet's core~\cite{Starkman:1990nj}. However, even in the case of a dense brown dwarf, and sub-GeV DM masses, we find that this ceiling is of the order of $\sigma_{\rm max} \sim 10^{-25}\, \rm cm^2$ (for the sub-GeV DM mass range).  Such cross section values are at the threshold where a point-like DM description is barely valid, and another physical description for DM must be used. Importantly, brown dwarfs provide complementary sensitivity to parameter space that can be tested by CR boosted DM, which can be difficult to interpret owing to no-energy dependence being used, despite being high-energy processes. The perhaps more robust bounds in this region (that are more clear to interpret when comparing with our regime) are those arising from Milky Way satelites~\cite{Nadler:2019zrb, Nadler:2020prv}, which are just above the range of cross sections shown in our plots.

\subsection{Discussion of Particle Physics Model Interpretations}

There are many classes of models to which our search has new sensitivity. In Sec~\ref{sec:equil}, we considered classes of models which have (i) $2 \rightarrow 2$ annihilation processes, and (ii) $3 \rightarrow 2$ annihilation processes. We provide some further details here, of specific realizations and general applicability.

\subsubsection{Details for $2 \rightarrow 2$ Annihilation Models}

We show that for $2 \rightarrow 2$ annihilation, the minimal required annihilation rate factor that can be probed with our search is $\langle \sigma v_{\rm rel} \rangle \gtrsim 10^{-37} - 10^{-34} \text{cm}^3/\rm s$. If we assume that DM is a thermal relic, it is generic in DM models that the annihilation rate proceeds either in the $s$-wave (which corresponds to no velocity suppression in the rate), or $p$-wave (velocity suppression). For annihilation, the size of $s$-wave rates today are $\sim10^{-26} \text{cm}^3/\rm s$, and $p$-wave rates are $\sim10^{-32} \text{cm}^3/\rm s$. These means that thermal $2\rightarrow2$ DM annihilation cross sections are generically testable with our search. As the small $p$-wave rate arises due to velocity suppression, rather than small couplings, it is not expected that the DM scattering rate is also suppressed for p-wave. The size of the elastic cross section is model dependent, however, for a concrete example see e.g. Ref.~\cite{Dutta:2019fxn}, where for a DM mass of 100 MeV the elastic nucleon cross section is $\sim10^{-35} \text{cm}^2$.

In terms of indirect detection constraints on these processes, $p$-wave annihilation cross sections are not even close to constrained by any existing experiments, see e.g. Refs.~\cite{Leane:2018kjk,Leane:2020liq,Cirelli:2020bpc}. For $s$-wave processes, there are CMB constraints for $\sim$ sub-GeV masses. However, the CMB bounds depend on details of the early Universe cosmology and are sensitive to variation of the annihilation rate with velocity. In addition, if the DM primarily interacts with leptons, both a freeze-in production~\cite{Knapen:2017xzo} and a WIMP-like $2\rightarrow 2$ freezeout production process~\cite{An:2017ojc} are experimentally allowed for sub-GeV masses. 

\subsubsection{Details for $3 \rightarrow 2$ Annihilation Models}

Strongly Interacting Massive Particles (SIMPs), generically have number changing freezeout processes, i.e. $3 \rightarrow 2$ annihilation. In known SIMP models, predicted elastic scattering cross sections are within our relevant parameter space. This is also true in Co-SIMP models, where the elastic scattering occurs at loop level, as shown in Ref.~\cite{Smirnov:2020zwf}.

To provide a detailed example, a DM particle thermally produced via the Co-SIMP mechanism would have an interaction rate factor of $\langle \sigma_{3 \rightarrow 2} v_{\rm rel}^2\rangle = 10^3 \left(m_\chi/\text{GeV}\right)^{-3} \rm GeV^{-5}$, while the annihilation equilibration condition requires $ \langle \sigma_{3 \rightarrow 2} v_{\rm rel}^2\rangle \geq  10^{-14} \left(m_\chi/\text{GeV}\right)^{-1}\, \rm GeV^{-5}$, which is satisfied by a thermal relic. The expected elastic scattering cross section induced at the loop level for e.g. a DM mass of 0.1 GeV is in the range of $\sigma \approx 10^{-32} \text{cm}^2 - 10^{-33} \text{cm}^2$~\cite{Smirnov:2020zwf}, and thus in the relevant parameter space we suggest to explore. This model is also not constrained by any indirect detection experiments.

\subsubsection{General Considerations}

More broadly, note that in order to contribute significantly to the heat flow, the DM population should have a dominant symmetric component (and not dominantly annihilate into invisible final states), which is a natural outcome in scenarios with thermally produced DM. Thermally produced DM candidates as discussed above, with sub-GeV masses are known to exist in models with light mediators~\cite{Knapen:2017xzo} and production mechanisms with number changing interactions~\cite{Hochberg:2014dra, Smirnov:2020zwf}.

Note that while we have cast our sensitivity in terms of one DM particle with one interaction type, in principle several particle processes may be present in the dark sector, which can alter the expected phenomenology (see e.g. Refs.~\cite{Kahlhoefer:2015bea,Bell:2016fqf,Bell:2016uhg,Duerr:2016tmh,Bell:2017irk,Cui:2017juz}). Detailed model-dependent studies would need to be performed to determine the full range of particle physics possibilities. Importantly, note that direct detection or other competing bounds may be weakened or removed in some DM models, while the exoplanet sensitivities would remain present. This can be true, for example, in inelastic DM models, which can evade direct detection limits, but could be probed with our search. Furthermore, since the exoplanet search features a probe of non-local DM densities, and direct detection assumes some local DM density, these are independent probes of the DM parameter space.

\section{Impact of Atmospheric Emissivity}
\label{app:emiss}

Exoplanet atmospheric emissivity can have an impact on JWST searches. This is because emissivity can trap some exoplanet heat flow, leading to higher temperatures. We now briefly demonstrate how this can improve JWST sensitivities.

\begin{figure*}[t!]
\centering
\includegraphics[width=0.45\columnwidth]{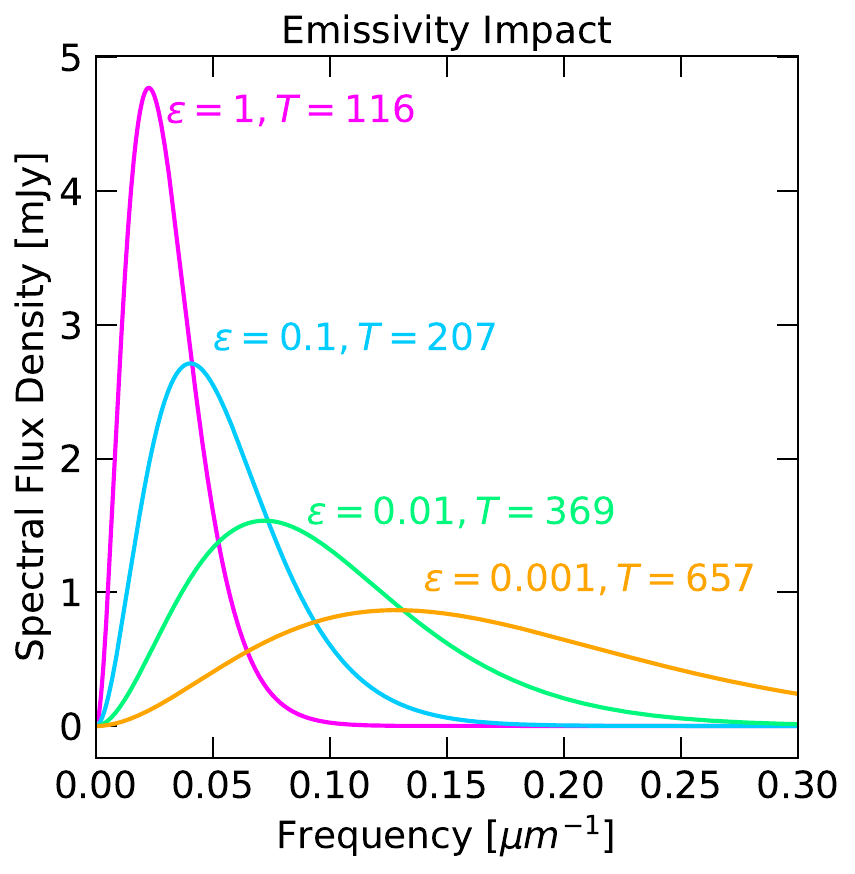}
\includegraphics[width=0.44\columnwidth]{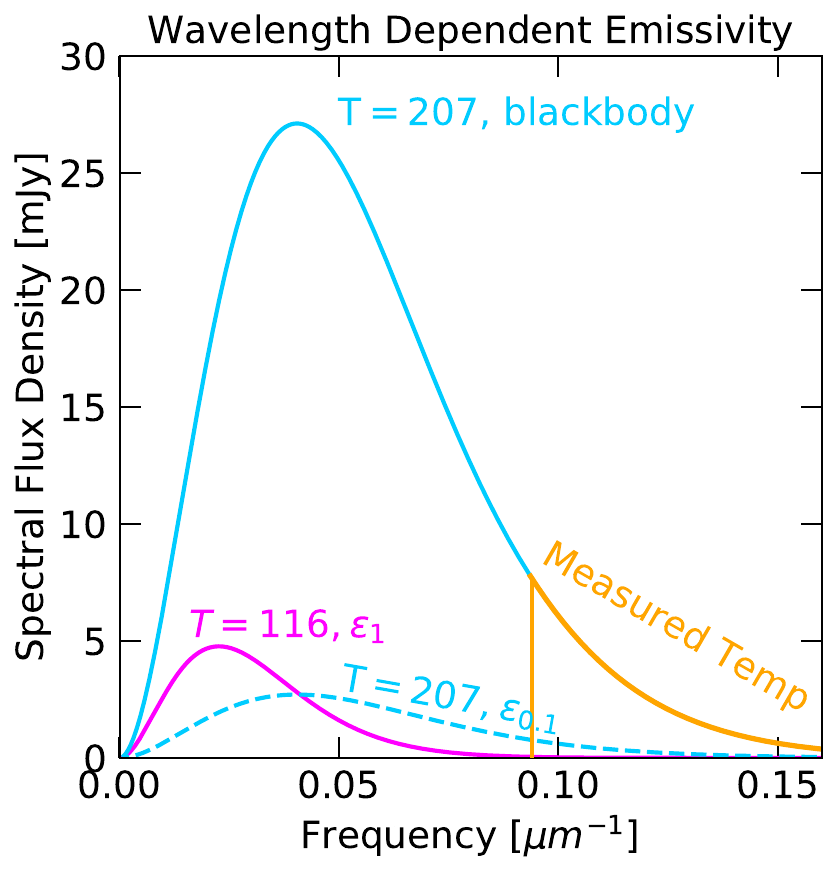}
\caption{Impact of atmospheric emissivity on heating signals. \textbf{Left:} Temperature peak shifts for smaller emissivities. \textbf{Right:} Schematic example of wavelength-dependent emissivity values, which can substantially boost signal intensity. This example has emissivity of one above about 0.08 $\mu m^{-1}$, and very suppressed emissivities below this frequency. This leads to a non-suppressed peak at higher frequencies, shown as the orange ``measured temp'' curve. The non-suppressed version of the emissivity equal to 0.1 curve is shown in solid blue. The dashed blue is the correctly rescaled version of the emissivity 0.1 case (i.e., the 0.1 penalty is applied to the blackbody temperature). The emissivity equal to one case is shown as solid magenta.}
\label{fig:emissiv}
\end{figure*}

Figure~\ref{fig:emissiv} (left) shows the impact of varied exoplanet emissivity on the spectral flux density. As the internal heat and the power output of a planet is a conserved quantity, a higher temperature can be obtained for smaller emissivity values, at the cost of a drop in the normalization of the spectral flux. As energy is conserved, this leads to the \textit{same} total integrated flux for all emissivities, but the temperature peaks at a shorter wavelength. The main benefit is therefore being able to exploit the more powerful filters available on JWST's instruments; longer wavelengths generally have worse flux sensitivity than the shorter wavelengths. The example scenario shown here is for a DM-heated Jupiter at 10 pc, which is a slice at $d=10~$pc through Fig.~\ref{fig:jov}. While an increased emissivity can lead to a larger effect also for the longer-distance searches, the effect is generally not as pronounced, as the higher temperature filters are not substantially more powerful than the already high-temperature filters used at large distances in the $\epsilon=1$ case shown in Fig.~2 of the main text.

Figure~\ref{fig:emissiv} (right) shows a schematic wavelength-dependent emissivity scenario. Indeed, in reality, an exoplanet may have different emissivities at different wavelengths, due to some wavelengths being better reflected by the atmosphere. For example, one could imagine an atmosphere leaving optical wavelengths mostly unaffected, while internally reflecting infrared wavelengths. This would lead to an extreme departure from the usual blackbody spectrum, similar to what is shown in the right figure. This also can allow both a boost in flux density compared to an emissivity value that is constant at all wavelengths, as well as applicability of better filters at shorter wavelengths. A planet could look, for example, truly like a higher temperature planet if only observing the edge of the spectrum (e.g. if a telescope was wavelength limited), without decreased normalization, while in other wavelengths, the normalization could be greatly suppressed due to the emissivity factor. In such a scenario, the area under the curves would still be conserved (e.g. the area under all of the smaller three fluxes in Fig.~\ref{fig:emissiv} is conserved). This variance of temperature peaks, at different emissivities, when the planet could not otherwise reach such high temperatures without DM, would be a smoking gun signal of a DM-heated planet. To be conservative, we do not use wavelength-dependent emissivities in our main results; we only point out this can potentially considerably boost sensitivities.

Lastly, note that at small emissivity values, the exoplanet surface temperature might become completely unaccessible, and DM-heating may instead only impact the temperature of the atmosphere in an energy exchange process. The details of such an effect will however depend on the exoplanet in question, and is an interesting possibility to study in a dedicated simulation, which is outside the scope of this work.

\bibliography{dwarfs}

\end{document}